\documentclass[11pt,twoside]{article}
\usepackage{mathrsfs}
\usepackage{amsmath}
\usepackage{amssymb}
\usepackage{graphicx}
\usepackage{CJKutf8}
\usepackage{subfigure}
\usepackage{appendix}
\usepackage{cite}
\numberwithin{equation}{section}

\newtheorem{proposition}{Proposition}

\newtheorem{corollary}{Corollary}
\newtheorem{remark}{Remark}

\DeclareMathOperator{\im}{Im}
\DeclareMathOperator{\re}{Re}

\newcommand*{\QEDB}{\hfill\ensuremath{\square}}

\title{On the complete integrability of space-time shifted nonlocal equations}
\author{Baoqiang Xia
\\
School of Mathematics and Statistics, Jiangsu Normal
University,\\
 Xuzhou, Jiangsu 221116, P. R. China,
 \\
 E-mail address:
xiabaoqiang@126.com
}

\date{}
\begin{document}
\begin{CJK}{UTF8}{gbsn}

\maketitle
\begin{abstract}

We investigate the complete integrability of soliton equations with shifted nonlocal reductions under the rapidly decreasing boundary conditions. The illustrative examples we choose are the Ablowitz-Ladik (AL) system and the Ablowitz-Kaup-Newell-Segur (AKNS) system. For this two models with the space and space-time shifted nonlocal reductions, we establish the complete integrability of the resulting nonlocal systems by an explicit construction of the variables of action-angle type from the corresponding scattering data. Moreover, we find that the time shifted nonlocal reductions, unlike the space and space-time shifted ones, are not compatible with the Poisson bracket relations of the corresponding scattering data in the presence of the discrete spectrum.


\end{abstract}
\newpage

\section{ Introduction}

Recently, it was found that soliton equations can support a new type of symmetry reductions involving a combination of a shift and a reflection \cite{AM2021,AMN2023,LWZ2022}. Such reductions are generalizations of the space or time nonlocal reductions introduced by Ablowitz and Mussiliman \cite{AM2013,AM2014}.
Since the work of Ablowitz and Mussilimani, various studies on the nonlocal equations have been developed. It is too vast to present faithfully all those studies here. For our purpose, we only refer, with apologies to missed authors, to \cite{AMN2023,AM20161,AM20162,F2016,C2018} and references therein for the topic of the implementations of the well-known inverse scattering transform (IST) to the nonlocal equations.

For the unreduced or standard (local) reduced systems, an important discovery is that the IST provides an infinite-dimensional analogue of Liouville theorem: the scattering data produce the canonical variables of action-angle type, confirming the complete integrability of the model in the infinite-dimensional Hamiltonian setting, see e.g. \cite{ZF1971,ZM1974,BFT1986,Faddeev2007}. In \cite{GS2017,G2019}, this result was extended to the space nonlocal reduction case.
A natural question is that if this programme can be further extended to the space, time and space-time shifted nonlocal equations. In this paper, we will show that it is indeed the case for the space and space-time shifted nonlocal equations, but it is not true for the time shifted nonlocal equations when the scattering data include  the discrete spectrum. We will choose the Ablowitz-Ladik (AL) system \cite{AL1,AL2,AL3} as our principal representative example. For this model with space-time shifted nonlocal reductions, we will present full details on the constructions of the variables of action-angle type from the corresponding scattering data. We will also briefly outline the analogous results for the corresponding continuous model, the well-known Ablowitz-Kaup-Newell-Segur (AKNS) system \cite{AKNS1974}.

The paper is organized as follows. In section 2, we review the main results on the inverse scattering problem associated with the AL system in order to introduce the required notations and tools.
In section 3, we derive the $r$-matrix representation of the Poisson brackets of the reduced monodromy matrix for the AL system. Such a representation enables us to compute all the Poisson bracket relations of the scattering data associated with the AL system. To the best of our knowledge, this result has not been reported in the literature. This is the main reason that we pay more attention to the AL system, rather than AKNS system (the use of $r$-matrix representation to derive the Poisson brackets of the scattering data for the AKNS system has been well exploited, see e.g. \cite{Faddeev2007}).
In section 4, we investigate the complete integrability of the AL system with shifted nonlocal reductions.
In particular, we establish the complete integrability of the space and space-time shifted nonlocal AL systems by an explicit construction of the variables of action-angle type, and we show that the time shifted nonlocal reduction is not compatible with the Poisson bracket relations of the scattering data in the presence of the discrete spectrum.
In section 5, we investigate the complete integrability of the space-time shifted nonlocal reductions in the context of the AKNS system.
Our results are summarized and discussed further in section 6.

\section{The AL system and the associated inverse scattering problem}

\subsection{The AL system and its reductions}

The AL lattice system \cite{AL1,AL2,AL3}
\begin{eqnarray}
\begin{array}{l}
iq_t(n,t)+q(n+1,t)-2q(n,t)+q(n-1,t)-p(n,t)q(n,t)\left( q(n+1,t)+ q(n-1,t)\right)=0,
\\
ip_t(n,t)-p(n+1,t)+2p(n,t)-q(n-1,t)+p(n,t)q(n,t)\left( p(n+1,t)+ p(n-1,t)\right)=0,
\end{array}
 \label{aleq1}
\end{eqnarray}
is an infinite-dimensional integrable Hamiltonian system described by the form \cite{GI1982,GIK1984,G2019,CC2019,Faddeev2007}
\begin{eqnarray}
q_t(n,t)=\left\{H,q(n,t)\right\},~~p_t(n,t)=\left\{H,p(n,t)\right\},
\label{HE}
\end{eqnarray}
with the Hamiltonian
\begin{eqnarray}
\begin{split}
H=-\sum_{n=-\infty}^{\infty}\left(q(n,t)(p(n-1,t)+p(n+1,t))+2\ln\left(1-q(n,t)p(n,t)\right)\right),
\end{split}
\label{H}
\end{eqnarray}
and the Poisson brackets
\begin{eqnarray}
\left\{q(n,t),q(m,t)\right\}=\left\{p(n,t),p(m,t)\right\}=0,
~~
\left\{q(n,t),p(m,t)\right\}=i\delta_{n,m}\left(1-q(n,t)p(n,t)\right).
\label{cPB}
\end{eqnarray}

This model admits rich reductions. The reductions mainly include the standard (local) reduction
\begin{eqnarray}
p(n,t)=\nu q^\ast(n,t), ~~\nu=\pm 1,
\label{red}
\end{eqnarray}
and the new space-time shifted nonlocal reductions \cite{AM2021,LWZ2022,AMN2023}
\begin{eqnarray}
p(n,t)=\nu q^\ast(n_0-n,t), ~~n_0\in \mathbb{Z},
\label{nred1}
\\
p(n,t)=\nu q(n,t_0-t), ~~t_0\in \mathbb{C},
\label{nred2}
\\
p(n,t)=\nu q(n_0-n,t_0-t), ~~n_0\in \mathbb{Z}[i], ~~t_0\in \mathbb{C},
\label{nred3}
\end{eqnarray}
where the asterisk denotes complex conjugation, and the notation $n_0\in \mathbb{Z}[i]$ means that the real and imaginary parts of the complex parameter $n_0$ are integers. Under these reductions, the AL system leads to the integrable discrete nonlinear Schr\"{o}dinger (DNLS) equation \cite{AL1,AL2,AL3}, the new space, time and space-time shifted nonlocal DNLS equations \cite{AM2021,AMN2023,LWZ2022}, respectively. They are
\begin{eqnarray}
iq_t(n,t)+q(n+1,t)-2q(n,t)+q(n-1,t)-\nu |q(n,t)|^2\left(q(n+1,t)+q(n-1,t)\right)=0.
\label{dnls}
\\
iq_t(n,t)=2q(n,t)-q(n+1,t)-q(n-1,t)+\nu q(n,t)q^\ast(n_0-n,t)\left(q(n+1,t)+q(n-1,t)\right),
\label{ssdnls}
\\
iq_t(n,t)=2q(n,t)-q(n+1,t)-q(n-1,t)+\nu q(n,t)q(n,t_0-t)\left(q(n+1,t)+q(n-1,t)\right),
\label{tsdnls}
\\
iq_t(n,t)=2q(n,t)-q(n+1,t)-q(n-1,t)+\nu q(n,t)q(n_0-n,t_0-t)\left(q(n+1,t)+q(n-1,t)\right).
\label{stsdnls}
\end{eqnarray}
We note that the space shifted nonlocal reduction (\ref{nred1}) and the resulting space shifted nonlocal DNLS equation (\ref{ssdnls}) were first found in \cite{Xia2019} (see equation (4.50) in \cite{Xia2019}).

\subsection{The inverse scattering problem for the AL system}
The aim of this subsection is to collect the main results on the inverse scattering problem for the AL system following \cite{AL1,AL2,AL3,XF2018}, and reformulate these results in a fashion that is convenient for deriving the $r$-matrix representation for the Poisson brackets of the corresponding scattering data. We will restrict our attention to the case with rapidly decreasing boundary conditions,
\begin{eqnarray}
\lim_{|n|\rightarrow \infty}p(n)=\lim_{|n|\rightarrow \infty} q(n)=0.
\end{eqnarray}

\subsubsection{The transition matrix and Jost solutions}

The starting point is that the AL system
(\ref{aleq1}) is equivalent to the compatibility condition
\begin{eqnarray}
L_t(n)=M(n+1)L(n)-L(n)M(n)
\label{zc}
\end{eqnarray}
of the following pair of linear spectral problems \cite{XF2018,MEKL1995,Geng2007}
\begin{subequations}
\begin{eqnarray}
\Phi(n+1,z)=L(n,z)\Phi(n,z),
\label{LPS}\\
\Phi_t(n,z)=M(n,z)\Phi(n,z),
\label{LPT}
\end{eqnarray}
\label{LPST}
\end{subequations}
where
\begin{eqnarray}
L(n,z)=\frac{1}{f(n)}\left(Z+Q(n)\right),
~~
M(n,z)=i\omega(z)\sigma_3+P(n,z),
\label{lpal}
\end{eqnarray}
with
\begin{eqnarray}
\begin{split}
&f(n)=\sqrt{1-q(n)p(n)},~~Z=\left( \begin{array}{cc} z & 0 \\
 0 &  z^{-1}\\ \end{array} \right), \quad
Q(n)=\left( \begin{array}{cc} 0 & q(n) \\
 p(n) &  0 \\ \end{array} \right),
 \\
&\omega(z)=\frac{1}{2} \left(z- z^{-1}\right)^2,~~\sigma_3=\left( \begin{array}{cc} 1 & 0 \\
 0 &  -1 \\ \end{array} \right),
 \\
 &P(n,z)=i\left( \begin{array}{cc} -\frac{1}{2} q(n)p(n-1)-\frac{1}{2} p(n)q(n-1)&  zq(n)- z^{-1}q(n-1) \\ zp(n-1)- z^{-1}p(n) & \frac{1}{2} q(n)p(n-1)+\frac{1}{2} p(n)q(n-1) \\ \end{array} \right).
\end{split}
\label{QH}
\end{eqnarray}
Here and in what follows, we have suppressed the $t$ dependence for conciseness unless there is ambiguity.
We note that here we use a different normalisation of the Lax pair compared to the traditional one introduced in \cite{AL1}. This normalized Lax pair has appealing algebraic properties \cite{MEKL1995,XF2018,CC2019}, which is more convenient for our analysis.

The transition matrix $T(n,m,z)$ is defined as the solution of (\ref{LPS}),
\begin{eqnarray}
T(n+1,m,z)=L(n,z)T(n,m,z),
\label{TL}
\end{eqnarray}
 with the initial condition
\begin{eqnarray}
\left.T(n,m,z)\right|_{n=m}=\mathbb{I},
\label{IT}
\end{eqnarray}
where $\mathbb{I}$ denotes the identity matrix.
For $n>m$, the transition matrix is given by
\begin{eqnarray}
T(n,m,z)=L(n-1,z)L(n-2,z)\cdots L(m,z),
\label{T1}
\end{eqnarray}
for $n<m$, it is given by
\begin{eqnarray}
T(n,m,z)=T^{-1}(m,n,z)=L^{-1}(n,z)L^{-1}(n+1,z)\cdots L^{-1}(m-1,z).
\label{T2}
\end{eqnarray}

The Jost solutions $T_{\pm}(n,z)$ for $|z|=1$ are defined by
\begin{eqnarray}
T_{\pm}(n,z)=\lim_{m\rightarrow \pm\infty}T(n,m,z)E(m,z),
\label{JS}
\end{eqnarray}
where
\begin{eqnarray}
E(m,z)=\left( \begin{array}{cc} z^m & 0 \\
 0 &  z^{-m}\\ \end{array} \right).
\end{eqnarray}
Alternatively, they can be defined as solutions of (\ref{LPS}) with the asymptotic conditions
\begin{eqnarray}
T_{\pm}(n,z)=E(n,z)+O(1),~~n\rightarrow \pm\infty.
\label{JSA}
\end{eqnarray}
The algebraic property $\det L(n,z)=1$ implies that the matrices $T_{\pm}(n,z)$ are unimodular.
Let $T^{(1)}_{\pm}(n,z)$ and $T^{(2)}_{\pm}(n,z)$ denote the first and second columns of the Jost solutions $T_{\pm}(n,z)$ respectively.
It can be shown (see e.g. \cite{AL3,XF2018}) that $T^{(1)}_{-}(n,z)$ and $T^{(2)}_{+}(n,z)$ can be analytically continued outside the unit circle, $|z|\geq 1$, whereas $T^{(2)}_{-}(n,z)$ and $T^{(1)}_{+}(n,z)$ can be analytically continued inside the unit circle, $|z|\leq 1$. Moreover, these columns have the following asymptotic behaviors \cite{XF2018}:
\begin{subequations}
\begin{eqnarray}
z^{-n}T^{(1)}_{-}(n,z)=\frac{C(n)}{\mathcal{C}}\left\{\left( \begin{array}{c} 1 \\
  0 \\ \end{array} \right)+\left( \begin{array}{c} O(z^{-2},\text{even}) \\
  O(z^{-1},\text{odd}) \\ \end{array} \right)\right\}, ~~ z\rightarrow \infty,
 \label{aspT1}
\\
 z^{n}T^{(2)}_{+}(n,z)=\frac{1}{C(n)}\left\{\left( \begin{array}{c} 0 \\
  1 \\ \end{array} \right)+\left( \begin{array}{c} O(z^{-1},\text{odd}) \\
  O(z^{-2},\text{even}) \\ \end{array} \right)\right\}, ~~ z\rightarrow \infty,
 \label{aspT2}
\\
z^{-n}T^{(1)}_{+}(n,z)=\frac{1}{C(n)}\left\{\left( \begin{array}{c} 1 \\
  0 \\ \end{array} \right)+\left( \begin{array}{c} O(z^{2},\text{even}) \\
  O(z,\text{odd}) \\ \end{array} \right)\right\}, ~~ z\rightarrow 0,
 \label{aspT3}
\\
z^{n}T^{(2)}_{-}(n,z)=\frac{C(n)}{\mathcal{C}}\left\{\left( \begin{array}{c} 0 \\
  1 \\ \end{array} \right)+\left( \begin{array}{c} O(z,\text{odd}) \\
  O(z^{2},\text{even}) \\ \end{array} \right)\right\}, ~~ z\rightarrow 0,
\label{aspT4}
\end{eqnarray}
\label{aspT}
\end{subequations}
where
\begin{eqnarray}
C(n)=\prod^{\infty}_{m=n}f(m)=\prod^{\infty}_{m=n}\sqrt{1-q(m)p(m)},
\quad
\mathcal{C}=\lim_{n\rightarrow-\infty}C(n).
\end{eqnarray}
In (\ref{aspT}), ``even" indicates that the higher-order terms are even powers of $z$ or $z^{-1}$, while ``odd" indicates that higher-order terms are odd powers.

\subsubsection{The scattering data}
The reduced monodromy matrix $T(z)$ is defined for $|z|=1$ as
\begin{eqnarray}
T(z)=T^{-1}_{+}(n,z)T_{-}(n,z).
\label{Tz}
\end{eqnarray}
It can be expressed as the limit
\begin{eqnarray}
T(z)=\lim\limits_{n\rightarrow \infty \atop m\rightarrow -\infty}E(-n,z)T(n,m,z)E(m,z)
=\lim_{n\rightarrow \infty}E(-n,z)T_{-}(n,z).
\label{Tzl}
\end{eqnarray}
Let us introduce the following notation for $T(z)$:
\begin{eqnarray}
T(z)=\left( \begin{array}{cc} a(z) & \tilde{b}(z) \\
 b(z) &  \tilde{a}(z)\\ \end{array} \right).
\end{eqnarray}
The unimodular property of the $n$-part of Lax pair (\ref{LPS}) implies that the matrix $T(z)$ is unimodular, i.e.
\begin{eqnarray}
a(z)\tilde{a}(z)-b(z)\tilde{b}(z)=1.
\label{abu}
\end{eqnarray}
By using (\ref{Tz}), we obtain the Wronskian representations for the elements of $T(z)$:
\begin{eqnarray}
\begin{split}
a(z)=\det\left(T^{(1)}_{-}(n,z),T^{(2)}_{+}(n,z)\right),
~~b(z)=-\det\left(T^{(1)}_{-}(n,z),T^{(1)}_{+}(n,z)\right),
\\
\tilde{a}(z)=-\det\left(T^{(2)}_{-}(n,z),T^{(1)}_{+}(n,z)\right),
~~\tilde{b}(z)=\det\left(T^{(2)}_{-}(n,z),T^{(2)}_{+}(n,z)\right).
\end{split}
\label{abexp}
\end{eqnarray}
The analytic properties of the Jost solutions imply that $a(z)$ and $\tilde{a}(z)$ have analytic continuation to the domains $|z|>1$ and $|z|<1$ respectively, whereas $b(z)$ and $\tilde{b}(z)$ in general can only be well-defined on the unit circle $|z|=1$. It follows from (\ref{aspT}) and (\ref{abexp}) that $a(z)$ and $\tilde{a}(z)$ have asymptotic behaviors
\begin{eqnarray}
\begin{split}
a(z)&=\frac{1}{\mathcal{C}}+O(z^{-2},even), \quad z\rightarrow \infty,
\\
\tilde{a}(z)&=\frac{1}{\mathcal{C}}+O(z^{2},even), \quad z\rightarrow 0.
\end{split}
\label{aasp}
\end{eqnarray}
Moreover, we have the following symmetry relations:
\begin{eqnarray}
a(-z)=a(z),\quad \tilde{a}(-z)=\tilde{a}(z), \quad b(-z)=-b(z),\quad \tilde{b}(-z)=-\tilde{b}(z).
\label{abs}
\end{eqnarray}

The symmetry relations (\ref{abs}) imply that the zeros of $a(z)$ and $\tilde{a}(z)$ always appear in pairs: if $z_j$ is a zero of $a(\lambda)$, then so does $-z_j$; if $\tilde{z}_l$ is a zero of $\tilde{a}(\lambda)$, then so does $-\tilde{z}_l$. To simplify our analysis we shall assume that $a(z)$ and $\tilde{a}(z)$ have only finite number of simple zeros in the corresponding analytic domains. Let us denote the set of the zeros of $a(z)$ by $\mathcal{Z}\equiv\left\{z_j\right\}_{j=1}^{N}\cup\left\{-z_j\right\}_{j=1}^{N}$ where $|z_j|>1$,
and the set of the zeros of $\tilde{a}(z)$ by  $\tilde{\mathcal{Z}}\equiv\left\{\tilde{z}_l\right\}_{l=1}^{\tilde{N}}\cup\left\{-\tilde{z}_l\right\}_{l=1}^{\tilde{N}}$ where $|\tilde{z}_l|<1$. The expression (\ref{abexp}) implies that for $z=z_j$, the column $T^{(1)}_{-}(n,z)$ is proportional to the column $T^{(2)}_{+}(n,z)$, and for $z=\tilde{z}_l$, the column $T^{(2)}_{-}(n,z)$ is proportional to the column $T^{(1)}_{+}(n,z)$.
Let $\gamma_j$ and $\tilde{\gamma}_l$ be the corresponding proportionality coefficients,
\begin{eqnarray}
\begin{split}
T^{(1)}_{-}(n,z_j)=\gamma_jT^{(2)}_{+}(n,z_j),~~j=1,\cdots,N,
\\
T^{(2)}_{-}(n,\tilde{z}_l)=\tilde{\gamma}_lT^{(1)}_{+}(n,\tilde{z}_l),~~l=1,\cdots,\tilde{N}.
\end{split}
\label{gam}
\end{eqnarray}
Let $\hat{\gamma}_j$ and $\hat{\tilde{\gamma}}_l$ be the proportionality coefficients corresponding to the paired zeros $-z_j$ and $-\tilde{z}_l$, respectively. We can deduce that $\hat{\gamma}_j=-\gamma_j$ and $\hat{\tilde{\gamma}}_l=-\tilde{\gamma}_l$.

Following the terminology used for the continuous models \cite{Faddeev2007}, we will call $a(z)$, $\tilde{a}(z)$, $b(z)$ and $\tilde{b}(z)$ transition coefficients for the continuous spectrum, and call $\gamma_j$ and $\tilde{\gamma}_l$ transition coefficients for the discrete spectrum.
The set
\begin{eqnarray}
\left\{b(z),~\tilde{b}(z),~\{z_j\}_{j=1}^N,~\{\tilde{z}_l\}_{l=1}^{\tilde{N}},~\{\gamma_j\}_{j=1}^N,
~\{\tilde{\gamma}_l\}_{l=1}^{\tilde{N}}\right\}
\label{sd}
\end{eqnarray}
constitutes the so-called scattering data for the AL system.

A very useful fact is that the transition coefficients $a(z)$ and $\tilde{a}(z)$ can be expressed in terms of their zeros and $b(z)\tilde{b}(z)$. Indeed, we have the following trace formulae:
\begin{eqnarray}
a(z)=\frac{1}{\mathcal{C}}\frac{z^{2\tilde{N}}\prod_{j=1}^N\left(z^2-z^2_j\right)}
{z^{2N}\prod_{j=1}^{\tilde{N}}\left(z^2-\tilde{z}^2_j\right)}\exp\left\{-\frac{1}{2\pi i}\oint_{|\xi|=1}\frac{\xi\ln\left(\xi^{2(N-\tilde{N})}(1+b(\xi)\tilde{b}(\xi))\right)}{\xi^2-z^2}d\xi\right\},
~|z|>1,
\label{aexp1}
\\
\tilde{a}(z)=\mathcal{C}\frac{\prod_{j=1}^{\tilde{N}}\left(z^2-\tilde{z}^2_j\right)}{\prod_{j=1}^N\left(z^2-z^2_j\right)}
\exp\left\{\frac{1}{2\pi i}\oint_{|\xi|=1}\frac{\xi\ln\left(\xi^{2(N-\tilde{N})}(1+b(\xi)\tilde{b}(\xi))\right)}{\xi^2-z^2}d\xi\right\},
~|z|<1.
\label{haexp1}
\end{eqnarray}
To derive the above two formulae, we define
\begin{eqnarray}
\hat{a}(z)=\mathcal{C}a(z)\frac{z^{2N}\prod_{j=1}^{\tilde{N}}\left(z^2-\tilde{z}^2_j\right)}
{z^{2\tilde{N}}\prod_{j=1}^N\left(z^2-z^2_j\right)},
~~
\hat{\tilde{a}}(z)=\mathcal{C}\tilde{a}(z)\frac{\prod_{j=1}^N\left(z^2-z^2_j\right)
\prod_{j=1}^{\tilde{N}}\left(-\tilde{z}^2_j\right)}
{\prod_{j=1}^{\tilde{N}}\left(z^2-\tilde{z}^2_j\right)\prod_{j=1}^N\left(-z^2_j\right)}.
\label{ahat}
\end{eqnarray}
Then it follows that the function $\hat{a}(z)$ is analytic for $|z|>1$ and it has no zeros, while $\hat{\tilde{a}}(z)$ is analytic for $|z|<1$ and it has no zeros. Moreover, they satisfy the asymptotic conditions
\begin{eqnarray}
\begin{split}
\hat{a}(z)&=1, \quad z\rightarrow \infty,
\\
\hat{\tilde{a}}(z)&=1, \quad z\rightarrow 0.
\end{split}
\label{hataasp}
\end{eqnarray}
By using the symmetry relations (\ref{abs}) and the asymptotic conditions (\ref{hataasp}), one may deduce that
\begin{eqnarray}
\frac{1}{2\pi i}\oint_{|\xi|=1}\frac{\xi\ln\hat{a}(\xi)}{\xi^2-z^2}d\xi
=\left\{ \begin{array}{l} -\ln\hat{a}(z),~~|z|>1, \\
  0,~~|z|<1, \\ \end{array} \right.
\label{hataexp}
\\
\frac{1}{2\pi i}\oint_{|\xi|=1}\frac{\xi\ln\hat{\tilde{a}}(\xi)}{\xi^2-z^2}d\xi
=\left\{ \begin{array}{l} 0,~~|z|>1, \\
  \ln\hat{\tilde{a}}(z),~~|z|<1. \\ \end{array} \right.
\label{hatatexp}
\end{eqnarray}
Adding the above two equations and using (\ref{ahat}) yields
\begin{eqnarray}
a(z)=\frac{1}{\mathcal{C}}\frac{z^{2\tilde{N}}\prod_{j=1}^N\left(z^2-z^2_j\right)}
{z^{2N}\prod_{j=1}^{\tilde{N}}\left(z^2-\tilde{z}^2_j\right)}\exp\left\{-\frac{1}{2\pi i}\oint_{|\xi|=1}\frac{\xi\ln\left(\hat{a}(\xi)\hat{\tilde{a}}(\xi)\right)}{\xi^2-z^2}d\xi\right\},
~~|z|>1,
\label{aexp}
\\
\tilde{a}(z)=\frac{1}{\mathcal{C}}
\frac{\prod_{j=1}^{\tilde{N}}\left(z^2-\tilde{z}^2_j\right)\prod_{j=1}^N\left(-z^2_j\right)}
{\prod_{j=1}^N\left(z^2-z^2_j\right)\prod_{j=1}^{\tilde{N}}\left(-\tilde{z}^2_j\right)}
\exp\left\{\frac{1}{2\pi i}\oint_{|\xi|=1}\frac{\xi\ln\left(\hat{a}(\xi)\hat{\tilde{a}}(\xi)\right)}{\xi^2-z^2}d\xi\right\},
~~|z|<1.
\label{haexp}
\end{eqnarray}
Equations (\ref{abu}) and (\ref{ahat}) yield, for $|z|=1$,
\begin{eqnarray}
\hat{a}(z)\hat{\tilde{a}}(z)=\left(1+b(z)\tilde{b}(z)\right)\mathcal{C}^2
\frac{z^{2N}\prod_{j=1}^{\tilde{N}}\left(-\tilde{z}^2_j\right)}{z^{2\tilde{N}}\prod_{j=1}^N\left(-z^2_j\right)}.
\label{ahata}
\end{eqnarray}
Substituting (\ref{ahata}) into (\ref{aexp}) and (\ref{haexp}), we obtain the expressions (\ref{aexp1}) and (\ref{haexp1}).

\begin{remark}
The trace formulae for the transition coefficients $a(z)$ and $\tilde{a}(z)$ reported in previous literatures (see e.g. \cite{GIK1984,AMN2023}) subject to the assumption that the number of the zeros of $a(z)$ is identical to that of $\tilde{a}(z)$. This assumption is reasonable for investigating the standard local reduction, since, under the standard local reduction, $a(z)$ and $\tilde{a}(z)$ subject to the symmetry $\tilde{a}(z)=a^{\ast}(\frac{1}{z^\ast})$ \cite{AL1,AL2,AL3}. However, for space and space-time shifted nonlocal reductions, there appear to be no extra symmetry relations between the transition coefficients $a(z)$ and $\tilde{a}(z)$ (see \cite{AMN2023} and the section 4 in this paper). Thus, it is necessary to derive a more general trace formulae without this assumption on the numbers of the zeros. Here we provide for the first time such trace formulae, see formulae (\ref{aexp1}) and (\ref{haexp1}).
\end{remark}

\subsubsection{Time evolution of the scattering data}

Using the zero curvature representation (\ref{zc}), we obtain the evolution equation of transition matrix $T(n,m,z)$:
\begin{eqnarray}
 \frac{d T(n,m,z)}{d t}=M(n,z)T(n,m,z)-T(n,m,z)M(m,z).
 \label{Tnmd}
\end{eqnarray}
Letting in (\ref{Tnmd}) $m\rightarrow\pm\infty$, we find that the Jost solutions satisfy the evolution equation
\begin{eqnarray}
\frac{dT_{\pm}(n,z)}{dt}=M(n,z)T_{\pm}(n,z)-i\omega(z)T_{\pm}(n,z)\sigma_3.
\label{Td}
\end{eqnarray}
Expression (\ref{Tzl}) together with equation (\ref{Td}) yield the evolution equations for the reduced monodromy matrix
\begin{eqnarray}
\frac{dT(z)}{dt}=i\omega(z)[\sigma_3,T(z)].
\label{Tt}
\end{eqnarray}
In components, (\ref{Tt}) becomes
\begin{eqnarray}
\frac{d a(z)}{dt}=0, ~~\frac{d \tilde{a}(z)}{dt}=0,
~~
\frac{d b(z)}{dt}=-2i\omega(z)b(z), ~~\frac{d \tilde{b}(z)}{dt}=2i\omega(z)\tilde{b}(z).
\label{abt}
\end{eqnarray}
The evolution equations of the transition coefficients $\gamma_j$ and $\tilde{\gamma}_l$ can be derived from (\ref{gam}) and (\ref{Td}), they are
\begin{eqnarray}
\begin{split}
\frac{d \gamma_j}{dt}=-2i\omega(z_j)\gamma_j,~~~~j=1,\cdots,N,
\\
\frac{d \tilde{\gamma}_l}{dt}=2i\omega(\tilde{z}_l)\tilde{\gamma}_l,~~l=1,\cdots,\tilde{N}.
\end{split}
\label{gamjt}
\end{eqnarray}

Equations (\ref{abt}) and (\ref{gamjt}) can easily be solved, so that we obtain the explicit time dependence of the transition coefficients
\begin{eqnarray}
\begin{split}
a(z,t)=a(z,0),~~\tilde{a}(z,t)=\tilde{a}(z,0),
\\
b(z,t)=\exp\left(-2i\omega(z)t\right)b(z,0),
~~
\tilde{b}(z,t)=\exp\left(2i\omega(z)t\right)\tilde{b}(z,0),
\\
z_j(t)=z_j(0),~~\gamma_j(t)=\exp\left(-2i\omega(z_j)t\right)\gamma_j(0),~~j=1,\cdots,N,
\\
\tilde{z}_j(t)=\tilde{z}_j(0),
~~\tilde{\gamma}_j(t)=\exp\left(2i\omega(\tilde{z}_j)t\right)\tilde{\gamma}_j(0),
~~j=1,\cdots,\tilde{N}.
\end{split}
\label{tct}
\end{eqnarray}

The inverse part of the scattering problem, namely constructing the potentials $q(n,t)$ and $p(n,t)$ from the known scattering data, can be accomplished by reformulating the AL scattering problem as an appropriate Riemann-Hilbert problem. For details on this issue, we refer the reader to references \cite{AL3,XF2018} for example.

\subsection{Conserved quantities and trace identities}

Equations (\ref{abt}) imply that $a(z)$ provides a generating function for the conserved quantities, and so does $\tilde{a}(z)$. By studying the expansion of $\ln a(z)$ in powers of $z^{-1}$ and the expansion of $\ln\tilde{a}(z)$ in powers of $z$, we can extract two sequences of conserved quantities for the AL system including its Hamiltonian. Details will be presented below.

We consider the expansion of the transition matrix $T(n,m,z)$:
\begin{eqnarray}
T(n,m,z)=\left(\mathbb{I}+W(n,z)\right)\exp\left(D(n,m,z)\right)
\left(\mathbb{I}+W(m,z)\right)^{-1},
\label{texp}
\end{eqnarray}
where $W$ is an off-diagonal matrix and $D$ is a diagonal matrix.
Inserting (\ref{texp}) into (\ref{TL}) and splitting the result into diagonal and off-diagonal parts, we obtain
\begin{eqnarray}
W(n+1,z)\exp\left(D(n+1,m,z)-D(n,m,z)\right)=\frac{1}{f(n)}\left(Q(n)+ZW(n,z)\right),
\label{odia}
\\
\exp\left(D(n+1,m,z)-D(n,m,z)\right)=\frac{1}{f(n)}\left(Z+Q(n)W(n,z)\right).
\label{dia}
\end{eqnarray}
By using (\ref{dia}) to eliminate $\exp \left(D(n,m,z)\right)$ from (\ref{odia}), we obtain
\begin{eqnarray}
W(n+1,z)Z-ZW(n,z)+W(n+1,z)Q(n)W(n,z)-Q(n)=0.
\label{W}
\end{eqnarray}
The difference equation (\ref{dia}) with the initial condition $\left.D(n,m,z)\right|_{n=m}=0$ implied by (\ref{IT}) can easily be solved,
\begin{eqnarray}
\exp\left(D(n,m,z)\right)=Z^{n-m}\prod_{k=m}^{n-1} \frac{1}{f(k)}\left(\mathbb{I}+Z^{-1}Q(k)W(k,z)\right).
\label{D}
\end{eqnarray}
We write the off-diagonal matrix $W(n,z)$ as
\begin{eqnarray}
W(n,z)=\left( \begin{array}{cc} 0 & W_{12}(n,z) \\
 W_{21}(n,z) &  0 \\ \end{array} \right).
\label{W1}
\end{eqnarray}
Substituting (\ref{W1}) together with the expansion
\begin{eqnarray}
W_{12}(n,z)=\sum_{j=1}^{\infty} W^{(j)}_{12}(n)z^{j},
~~
W_{21}(n,z)=\sum_{j=1}^{\infty} W^{(j)}_{21}(n)z^{-j},
\label{Wexp}
\end{eqnarray}
into (\ref{W}), we obtain
\begin{eqnarray}
W^{(2j)}_{12}(n)=W^{(2j)}_{21}(n)=0,~~j\geq 1,
\label{W2j}
\end{eqnarray}
and $W^{(2j+1)}_{12}(n)$ and $W^{(2j+1)}_{21}(n)$ are given recursively by
\begin{eqnarray}
\begin{split}
W^{(1)}_{12}(n)=q(n-1), ~~W^{(1)}_{21}(n)=p(n-1),
\\
W^{(2j+1)}_{12}(n)=W^{(2j-1)}_{12}(n-1)-p(n-1)\sum\limits_{k+l=2j, \atop k,l\geq 1}W^{(k)}_{12}(n-1)W^{(l)}_{12}(n),~~j\geq 1,
\\
W^{(2j+1)}_{21}(n)=W^{(2j-1)}_{21}(n-1)-q(n-1)\sum\limits_{k+l=2j, \atop k,l\geq 1}W^{(k)}_{21}(n-1)W^{(l)}_{21}(n),~~j\geq 1.
\end{split}
\label{W2j}
\end{eqnarray}

By using (\ref{Tzl}), (\ref{texp}) and (\ref{D}), we find
\begin{eqnarray}
\ln a(z)=-\sum_{n=-\infty}^{\infty}\ln\left(f(n)\right)
+\sum_{n=-\infty}^{\infty}\ln\left(1+z^{-1}q(n)W_{21}(n,z)\right),
\label{lnaexp1}
\\
\ln \tilde{a}(z)=-\sum_{n=-\infty}^{\infty}\ln\left(f(n)\right)
+\sum_{n=-\infty}^{\infty}\ln\left(1+zp(n)W_{12}(n,z)\right).
\label{lnaexp2}
\end{eqnarray}
By computing the expansion of the logarithmic function $\ln\left(1+z^{-1}q(n)W_{21}(n,z)\right)$ in $z^{-1}$,
we obtain the asymptotic representation for $\ln a(z)$,
\begin{eqnarray}
\ln a(z)=I_0+\sum_{k=1}^{\infty}I_{2k} z^{-2k},
\label{lnaexp3}
\end{eqnarray}
where
\begin{eqnarray}
\begin{split}
I_0=-\sum_{n=-\infty}^{\infty}\ln\left(f(n)\right),
~~
I_2=\sum_{n=-\infty}^{\infty}q(n)p(n-1),
\\
I_{2k}=\sum_{n=-\infty}^{\infty}\sum_{m=1}^k\frac{(-1)^{m-1}}{m}(q(n))^m\sum_{k_1+\cdots+k_m=k}
W^{(2k_1-1)}_{21}(n)W^{(2k_2-1)}_{21}(n)\cdots W^{(2k_m-1)}_{21}(n).
\end{split}
\label{Ik}
\end{eqnarray}
Similarly, by computing the expansion of the logarithmic function $\ln\left(1+zp(n)W_{12}(n,z)\right)$ in $z$,
we obtain the asymptotic representation for $\ln \tilde{a}(z)$,
\begin{eqnarray}
\ln \tilde{a}(z)=\tilde{I}_0+\sum_{k=1}^{\infty}\tilde{I}_{2k} z^{2k},
\label{lnaexp4}
\end{eqnarray}
where
\begin{eqnarray}
\begin{split}
\tilde{I}_0=I_0=-\sum_{n=-\infty}^{\infty}\ln\left(f(n)\right),
~~
\tilde{I}_2=\sum_{n=-\infty}^{\infty}p(n)q(n-1),
\\
\tilde{I}_{2k}=\sum_{n=-\infty}^{\infty}\sum_{m=1}^k\frac{(-1)^{m-1}}{m}(p(n))^m\sum_{k_1+\cdots+k_m=k}
W^{(2k_1-1)}_{12}(n)W^{(2k_2-1)}_{12}(n)\cdots W^{(2k_m-1)}_{12}(n).
\end{split}
\label{Ik2}
\end{eqnarray}

The Poisson commutativity of the two sequences of the conserved quantities will be proved in the next section; see corollary 1 section 3.

On the other hand, $\ln a(z)$ and $\ln \tilde{a}(z)$ can be represented in terms of the scattering data as
\begin{eqnarray}
\ln a(z)=c_0+\sum_{k=1}^{\infty}c_{2k}z^{-2k},
\label{lnas}
\\
\ln \tilde{a}(z)=\tilde{c}_0+\sum_{k=1}^{\infty}\tilde{c}_{2k}z^{2k},
\label{lnhas}
\end{eqnarray}
where
\begin{eqnarray}
\begin{split}
c_0=-\ln \mathcal{C},
~~
\tilde{c}_0=\ln \mathcal{C}
+\ln\frac{\prod_{j=1}^{\tilde{N}}(-\tilde{z}^2_j)}{\prod_{j=1}^N(-z^2_j)}
+2(N-\tilde{N})\pi i
+\frac{1}{2\pi i}\oint_{|\xi|=1}\frac{1}{\xi}\ln(1+b(\xi)\tilde{b}(\xi))d\xi,
\\
c_{2k}=\frac{1}{k}\left(N-\tilde{N}+\sum_{j=1}^{\tilde{N}}\tilde{z}^{2k}_j-\sum_{j=1}^N z^{2k}_j\right)+\frac{1}{2\pi i}\oint_{|\xi|=1}\xi^{2k-1}\ln(1+b(\xi)\tilde{b}(\xi))d\xi,~k\geq 1,
\\
\tilde{c}_{2k}=\frac{1}{k}\left(\tilde{N}-N+\sum_{j=1}^Nz^{-2k}_j-\sum_{j=1}^{\tilde{N}}\tilde{z}^{-2k}_j\right)
+\frac{1}{2\pi i}\oint_{|\xi|=1}\xi^{-2k-1}\ln(1+b(\xi)\tilde{b}(\xi))d\xi, ~k\geq 1.
\end{split}
\label{cj}
\end{eqnarray}
Indeed, (\ref{lnas}) and (\ref{lnhas}) can be derived by computing the expansion of (\ref{aexp1}) in $z^{-1}$ and the expansion of (\ref{haexp1}) in $z$, and using
\begin{eqnarray}
\oint_{|\xi|=1}\xi^k\ln\xi^{2(N-\tilde{N})} d\xi
=\left\{ \begin{array}{l} -4(N-\tilde{N})\pi^2,~~k=-1, \\
  \frac{4(N-\tilde{N})\pi i}{k+1},~~k\neq -1, ~~k\in \mathbb{Z}, \\ \end{array} \right.
\end{eqnarray}
which can be verified easily by setting $\xi=e^{i\theta}$, $0\leq\theta\leq 2\pi$, in the integrand.

By comparing the asymptotic expansions (\ref{lnaexp3}) and (\ref{lnaexp4}) with expansions (\ref{lnas}) and (\ref{lnhas}), we obtain the trace identities for the AL system,
\begin{eqnarray}
\begin{split}
I_0=\tilde{I}_0=\frac{1}{2}\ln\frac{\prod_{j=1}^{\tilde{N}}\left(-\tilde{z}^2_j\right)}{\prod_{j=1}^N\left(-z^2_j\right)}
+(N-\tilde{N})\pi i
+\frac{1}{4\pi i}\oint_{|\xi|=1}\frac{1}{\xi}\ln\left(1+b(\xi)\tilde{b}(\xi)\right)d\xi,
\\
I_{2k}=c_{2k},~~\tilde{I}_{2k}=\tilde{c}_{2k},~~k\geq 1.
\end{split}
\label{Icj}
\end{eqnarray}

We note that the Hamiltonian of the AL system can be recovered from the combination of the first two nontrivial conserved quantities constructed above.
Indeed, the Hamiltonian (\ref{H}) can be recovered as
\begin{eqnarray}
H=4I_0-I_2-\tilde{I}_2.
\label{HI}
\end{eqnarray}
This confirms further the advantage of the use of the normalized Lax pair (\ref{LPST})
(if we prefer to the traditional one, the conserved quantity $I_0$ will be missed in the sequence of the conserved quantities obtained from the expansion of the corresponding transition coefficient $a(z)$).
With the observation (\ref{HI}) and the trace identities (\ref{Icj}), we are able to derive the representation of the Hamiltonian (\ref{H}) in terms of the scattering data:
\begin{eqnarray}
\begin{split}
H=&\sum_{j=1}^{\tilde{N}}\ln\tilde{z}_j^4-\sum_{j=1}^N\ln z^4_j
+4(N-\tilde{N})\pi i+\sum_{j=1}^N\left(z^{2}_j-z^{-2}_j\right)
-\sum_{j=1}^{\tilde{N}}\left(\tilde{z}^{2}_j-\tilde{z}^{-2}_j\right)
\\
&+\frac{1}{2\pi i}\oint_{|\xi|=1}\left(2\xi^{-1}-\xi^{-3}-\xi\right)\ln\left(1+b(\xi)\tilde{b}(\xi)\right)d\xi.
\end{split}
\label{Hs}
\end{eqnarray}

\section{Poisson structure on the scattering data for the AL system}

The aim of this section is to derive the Poisson structure on the scattering data of the AL system from the initial Poisson structure (\ref{cPB}). Different from the method based on expansions over the squared solutions \cite{GIK1984,G2019}, our analysis is based on the tool of $r$-matrix \cite{Faddeev2007}. In particular, we shall derive the $r$-matrix representation of the Poisson brackets of the reduced monodromy matrix for the AL system. To the best of our knowledge, this is the first time such a result is derived and it is the first time that the Poisson bracket relations of the scattering data for the AL system is derived from a $r$-matrix representation. A byproduct is that the conserved quantities introduced in section 2.4 are proved to be in involution.

Using the initial Poisson brackets (\ref{cPB}), one can check directly that the Lax matrix $L(n,z)$ defined by (\ref{lpal}) satisfies the following ultralocal Poisson algebra \cite{CC2019}
\begin{eqnarray}
\left\{L(n,z)\underset{,}{\otimes}L(m,\mu)\right\}=\delta_{n,m}\left[r(\frac{z}{\mu}),L(n,z)\otimes L(m,\mu)\right],
\label{LPB}
\end{eqnarray}
where
\begin{eqnarray}
r(z)=
\frac{i}{2(1-z^2)}\left( \begin{array}{cccc}
z^2+1 & 0 & 0 & 0
\\
0 &  0 & 2z & 0
\\
0 & 2z & 0 & 0
\\
0 &  0 & 0 & z^2+1
 \\ \end{array} \right).
\label{rm1}
\end{eqnarray}
For (\ref{rm1}), it is straightforward to verify that the relation
\begin{eqnarray}
r(z)\left(A\otimes B\right)=\left(B\otimes A\right)r(z)
\label{rab}
\end{eqnarray}
holds for any two $2\times 2$ diagonal matrices $A$ and $B$.

We begin by deriving the Poisson brackets of the Jost solutions.
\begin{proposition}\label{PBJ}
The Poisson brackets of the Jost solutions $T_{\pm}(n,z)$ for $|z|=1$ are given by
\begin{eqnarray}
\begin{split}
\left\{T_{-}(n,z)\underset{,}{\otimes}T_{-}(n,\mu)\right\}=
&r(\frac{z}{\mu})\left(T_{-}(n,z)\otimes T_{-}(n,\mu)\right)-\left(T_{-}(n,z)\otimes T_{-}(n,\mu)\right)r_{-}(z,\mu),
\\
\left\{T_{+}(n,z)\underset{,}{\otimes}T_{+}(n,\mu)\right\}=
&\left(T_{+}(n,z)\otimes T_{+}(n,\mu)\right)r_{+}(z,\mu)-r(\frac{z}{\mu})\left(T_{+}(n,z)\otimes T_{+}(n,\mu)\right),
\\
\left\{T_{-}(n,z)\underset{,}{\otimes}T_{+}(n,\mu)\right\}=&0,
\end{split}
\label{JPB}
\end{eqnarray}
where
\begin{eqnarray}
\begin{split}
r_{\pm}(z,\mu)=&\lim_{m\rightarrow\pm\infty}\left(E(m,\frac{\mu}{z})\otimes E(m,\frac{z}{\mu})\right)r(\frac{z}{\mu})
\\=&\frac{i}{2}\left( \begin{array}{cccc}
p.v.\frac{\mu^2+z^2}{\mu^2-z^2} & 0 & 0 & 0
\\
0 &  0 & \pm\pi i\left(\delta(\frac{z}{\mu})-\delta(-\frac{z}{\mu})\right) & 0
\\
0 & \mp\pi i\left(\delta(\frac{z}{\mu})-\delta(-\frac{z}{\mu})\right) & 0 & 0
\\
0 &  0 & 0 & p.v.\frac{\mu^2+z^2}{\mu^2-z^2}
 \\ \end{array} \right).
\end{split}
\label{rm}
\end{eqnarray}
Here p.v. indicates principal value and the delta function $\delta(\frac{z}{\mu})$ is defined by
\begin{eqnarray}
\int_{|\mu|=1}\delta(\pm\frac{z}{\mu})\frac{f(\mu)}{\mu}d\mu=f(\pm z).
\end{eqnarray}
\end{proposition}

{\bf Proof}
Using (\ref{T1}), (\ref{T2}) and (\ref{LPB}), we obtain that the entries of the transition matrix $T(n,m,z)$ satisfy the Poisson brackets in the form
\begin{eqnarray}
\begin{split}
\left\{T(n,m,z)\underset{,}{\otimes}T(n,m,\mu)\right\}=\left[r(\frac{z}{\mu}),T(n,m,z)\otimes T(n,m,\mu)\right], ~~m<n,
\\
\left\{T(n,m,z)\underset{,}{\otimes}T(n,m,\mu)\right\}=-\left[r(\frac{z}{\mu}),T(n,m,z)\otimes T(n,m,\mu)\right], ~~m>n,
\\
\left\{T(n,m_1,z)\underset{,}{\otimes}T(n,m_2,\mu)\right\}=0,~~m_1<n<m_2.
\end{split}
\label{TPB}
\end{eqnarray}
The Poisson algebra (\ref{JPB}) can be derived by using (\ref{rab}), (\ref{TPB}) and the definition (\ref{JS}), and the following formula \cite{Faddeev2007}
\begin{eqnarray}
\lim_{n\rightarrow \pm\infty} p.v.\frac{\left(z\mu^{-1}\right)^n}{1-z\mu^{-1}}=\mp i\pi\delta(\frac{z}{\mu}),
\end{eqnarray}
for $|z|=|\mu|=1$. \QEDB

Using (\ref{Tzl}) and (\ref{JPB}), we obtain the following important relation.
\begin{proposition}\label{PBT}
For $|z|=|\mu|=1$, the Poisson brackets of the reduced monodromy matrix are given by
\begin{eqnarray}
\left\{T(z)\underset{,}{\otimes}T(\mu)\right\}=
r_{+}(z,\mu)\left(T(z)\otimes T(\mu)\right)-\left(T(z)\otimes T(\mu)\right)r_{-}(z,\mu),
\label{rTPB}
\end{eqnarray}
where $r_{\pm}(z,\mu)$ is given by (\ref{rm}).
\end{proposition}

The components of the matrix Poisson brackets (\ref{rTPB}) read:
\begin{subequations}
\begin{eqnarray}
\{a(z),a(\mu)\}=\{a(z),\tilde{a}(\mu)\}=\{\tilde{a}(z),\tilde{a}(\mu)\}
=0,
\label{aa}
\\
\{a(z),\tilde{b}(\mu)\}=-i\left(\frac{1}{ze^{+0}-u}+\frac{1}{ze^{+0}+u}\right)\frac{z^2+\mu^2}{4z}a(z)\tilde{b}(\mu),
\label{abtilde}
\\
\{a(z),b(\mu)\}=i\left(\frac{1}{ze^{+0}-u}+\frac{1}{ze^{+0}+u}\right)\frac{z^2+\mu^2}{4z}a(z)b(\mu),
\label{ab}
\\
\{\tilde{a}(z),\tilde{b}(\mu)\}=i\left(\frac{1}{ze^{-0}-u}+\frac{1}{ze^{-0}+u}\right)\frac{z^2+\mu^2}{4z}\tilde{a}(z)\tilde{b}(\mu),
\label{atildebtiled}\\
\{\tilde{a}(z),b(\mu)\}=-i\left(\frac{1}{ze^{-0}-u}+\frac{1}{ze^{-0}+u}\right)\frac{z^2+\mu^2}{4z}\tilde{a}(z)b(\mu),
\label{atildeb}
\\
\{b(z),b(\mu)\}=\{\tilde{b}(z),\tilde{b}(\mu)\}=0,~\{b(z),\tilde{b}(\mu)\}=\pi\left(\delta(\frac{z}{\mu})-\delta(-\frac{z}{\mu})\right)a(z)\tilde{a}(\mu).
\label{bb}
\end{eqnarray}
\label{abPB}
\end{subequations}
Here we have used the Sochocki-Plemelj formula
\begin{eqnarray}
\begin{split}
\frac{1}{ze^{-0}\pm u}=\lim\limits_{\tilde{z}\rightarrow z \atop |\tilde{z}|<1}\frac{1}{\tilde{z}\pm\mu}=p.v.\frac{1}{z\pm\mu}- \pi i\delta(\mp\frac{z}{\mu}),
\\
\frac{1}{ze^{+0}\pm u}=\lim\limits_{\tilde{z}\rightarrow z \atop |\tilde{z}|>1}\frac{1}{\tilde{z}\pm\mu}=p.v.\frac{1}{z\pm\mu}+\pi i\delta(\mp\frac{z}{\mu}),
\end{split}
\label{SPF}
\end{eqnarray}
in order to derive (\ref{abPB}).
Due to the analytic properties of $a(z)$ and $\tilde{a}(z)$, the Poisson brackets involving $a(z)$ in (\ref{abPB}) can be analytically continued into the domain $|z|>1$, while the ones involving $\tilde{a}(z)$ can be analytically continued into the domain $|z|<1$.

From (\ref{aa}), we immediately obtain
\begin{corollary}
The conserved quantities $I_{2k}$ and $\tilde{I}_{2k}$ constructed in section 2.4 (see (\ref{Ik}) and (\ref{Ik2})) are in involution,
\begin{eqnarray}
\left\{I_{2j},I_{2k}\right\}=\left\{\tilde{I}_{2j},\tilde{I}_{2k}\right\}
=\left\{I_{2j},\tilde{I}_{2k}\right\}=0,~~j,k\geq 0.
\end{eqnarray}
\end{corollary}

With aid of (\ref{JPB}) and (\ref{abPB}), we are able to compute all the Poisson bracket relations between the scattering data.
\begin{proposition}\label{disPoisson}
For the AL system, the Poisson brackets of the scattering data (\ref{sd}) are given by
\begin{subequations}
\begin{eqnarray}
\left\{z_j,z_k\right\}=\left\{z_j,\tilde{z}_l\right\}=\left\{\tilde{z}_l,\tilde{z}_m\right\}=0,~~1\leq j,k\leq N,~~1\leq l,m\leq \tilde{N},
\label{zjk}
\\
\left\{\gamma_j,\gamma_k\right\}=\left\{\gamma_j,\tilde{\gamma}_l\right\}=\left\{\tilde{\gamma}_l,\tilde{\gamma}_m\right\}=0,~~1\leq j,k\leq N,~~1\leq l,m\leq \tilde{N},
\label{gamjk}
\\
\left\{b(z),z_j\right\}=\left\{\tilde{b}(z),z_j\right\}=\left\{b(z),\tilde{z}_l\right\}
=\left\{\tilde{b}(z),\tilde{z}_l\right\}=0,~~1\leq j\leq N,~~1\leq l\leq \tilde{N},
\label{bzjk}
\\
\left\{b(z),\gamma_j\right\}=\left\{\tilde{b}(z),\gamma_j\right\}=\left\{b(z),\tilde{\gamma}_l\right\}
=\left\{\tilde{b}(z),\tilde{\gamma}_l\right\}=0,~~1\leq j\leq N,~~1\leq l\leq \tilde{N},
\label{bgamjk}
\\
\left\{z_j,\tilde{\gamma}_l\right\}=0,~~\left\{\gamma_j,z_k\right\}=\frac{i}{2}\delta_{jk}\gamma_jz_k,~~1\leq j,k\leq N,~~1\leq l\leq \tilde{N},
\label{zgamtilde}
\\
\left\{\tilde{z}_l,\gamma_j\right\}=0,
~~\left\{\tilde{\gamma}_l,\tilde{z}_m\right\}=\frac{i}{2}\delta_{lm}\tilde{\gamma}_l\tilde{z}_m
~~1\leq j\leq N,~~1\leq l,m\leq \tilde{N},
\label{gamz}
\\
\{b(z),b(\mu)\}=\{\tilde{b}(z),\tilde{b}(\mu)\}=0,
~\{b(z),\tilde{b}(\mu)\}=\pi\left(\delta(\frac{z}{\mu})-\delta(-\frac{z}{\mu})\right)\left(1+b(z)\tilde{b}(z)\right).
\label{bb1}
\end{eqnarray}
\label{abzpb}
\end{subequations}
\end{proposition}
{\bf Proof} The Poisson brackets (\ref{zjk}) follow from (\ref{aa}). The Poisson brackets (\ref{bb1}) follow from (\ref{abu}), (\ref{abs}) and (\ref{bb}).
In order to derive Poisson brackets involving $\gamma_j$ and $\tilde{\gamma}_l$, we introduce the following notations
\begin{eqnarray}
T_{-}(n,z)=\left( \begin{array}{cc} f_{-}(n,z) & \hat{g}_{-}(n,z) \\
 g_{-}(n,z) & \hat{f}_{-}(n,z) \\ \end{array} \right),
 ~~
 T_{+}(n,z)=\left( \begin{array}{cc} \hat{g}_{+}(n,z) & f_{+}(n,z) \\
 \hat{f}_{+}(n,z) & g_{+}(n,z)  \\ \end{array} \right).
\label{FG}
\end{eqnarray}
With the notation (\ref{FG}), the quantities $a(z)$, $b(z)$, $\tilde{a}(z)$, $\tilde{b}(z)$, $\gamma_j$ and $\tilde{\gamma}_l$ can be expressed as
\begin{eqnarray}
\begin{split}
a(z)=g_{+}(n,z)f_{-}(n,z)-f_{+}(n,z)g_{-}(n,z),
~~
b(z)=\hat{g}_{+}(n,z)g_{-}(n,z)-\hat{f}_{+}(n,z)f_{-}(n,z),
\\
\tilde{a}(z)=\hat{g}_{+}(n,z)\hat{f}_{-}(n,z)-\hat{f}_{+}(n,z)\hat{g}_{-}(n,z),
~~
\tilde{b}(z)=\hat{g}_{-}(n,z)g_{+}(n,z)-\hat{f}_{-}(n,z)f_{+}(n,z),
\\
\gamma_j=\left.\frac{f_{-}(n,z)}{f_{+}(n,z)}\right|_{z=z_j}
=\left.\frac{g_{-}(n,z)}{g_{+}(n,z)}\right|_{z=z_j},
~~\tilde{\gamma}_l=\left.\frac{\hat{f}_{-}(n,z)}{\hat{f}_{+}(n,z)}\right|_{z=\tilde{z}_l}
=\left.\frac{\hat{g}_{-}(n,z)}{\hat{g}_{+}(n,z)}\right|_{z=\tilde{z}_l}.
\end{split}
\label{abgamexp}
\end{eqnarray}
After straightforward calculations using (\ref{JPB}) and the expressions (\ref{abgamexp}), we obtain (\ref{gamjk}), (\ref{bgamjk}) and
\begin{subequations}
\begin{eqnarray}
\left\{a(z),\gamma_j\right\}=\frac{i}{2}\frac{z^2+z_j^2}{z^2-z_j^2}a(z)\gamma_j,
~~
\left\{a(z),\tilde{\gamma}_l\right\}=-\frac{i}{2}\frac{z^2+\tilde{z}_l^2}{z^2-\tilde{z}_l^2}a(z)\tilde{\gamma}_l,
\label{agampb}
\\
\left\{\tilde{a}(z),\gamma_j\right\}=-\frac{i}{2}\frac{z^2+z_j^2}{z^2-z_j^2}\tilde{a}(z)\gamma_j,
~~
\left\{\tilde{a}(z),\tilde{\gamma}_l\right\}=\frac{i}{2}\frac{z^2+\tilde{z}_l^2}{z^2-\tilde{z}_l^2}\tilde{a}(z)\tilde{\gamma}_l.
\label{tagampb}
\end{eqnarray}
\end{subequations}
From (\ref{ahat}) and the first of (\ref{agampb}), we obtain
\begin{eqnarray}
\{\ln \frac{z^{2\tilde{N}}\hat{a}(z)}{z^{2N}\mathcal{C}},\gamma_j\}
+\sum_{k=1}^N(\frac{\{z_k,\gamma_j\}}{z+z_k}-\frac{\{z_k,\gamma_j\}}{z-z_k})
+\sum_{l=1}^{\tilde{N}}(\frac{\{\tilde{z}_l,\gamma_j\}}{z-\tilde{z}_l}
-\frac{\{\tilde{z}_l,\gamma_j\}}{z+\tilde{z}_l})
=\frac{i(z^2+z_j^2)}{2(z^2-z_j^2)}\gamma_j.
\label{tagampb1}
\end{eqnarray}
For $ |z|>1$, the right hand side of equality (\ref{tagampb1}) has singularities at $z=\pm z_j$.
Comparing the residues at $z=\pm z_j$, we find the second of relations (\ref{zgamtilde}).
From (\ref{ahat}) and the second of (\ref{agampb}), we obtain
\begin{eqnarray}
\{\ln \frac{z^{2\tilde{N}}\hat{a}(z)}{z^{2N}\mathcal{C}},\tilde{\gamma}_l\}
+\sum_{k=1}^N(\frac{\{z_k,\tilde{\gamma}_l\}}{z+z_k}-\frac{\{z_k,\tilde{\gamma}_l\}}{z-z_k})
+\sum_{m=1}^{\tilde{N}}(\frac{\{\tilde{z}_m,\tilde{\gamma}_l\}}{z-\tilde{z}_m}
-\frac{\{\tilde{z}_m,\tilde{\gamma}_l\}}{z+\tilde{z}_m})
=\frac{i(z^2+\tilde{z}_l^2)}{2(\tilde{z}_l^2-z^2)}\tilde{\gamma}_l.
\label{tagampb2}
\end{eqnarray}
Since the right hand side of (\ref{tagampb2}) is analytic for $|z|>1$, the left hand side has no singularities at $z=\pm z_j$. Thus we obtain the first of relations (\ref{zgamtilde}). The Poisson brackets (\ref{gamz}) can be derived from (\ref{tagampb}) via a similar manner. The Poisson brackets (\ref{bzjk}) can be deduced by using (\ref{ahat}) and (\ref{abPB}).
\QEDB

The Poisson brackets (\ref{abzpb}) suggest that the quantities
\begin{eqnarray}
\begin{split}
\rho\left(z\right)=\frac{1}{\pi}\ln\left(1+b(z)\tilde{b}(z)\right),
~~\phi\left(z\right)=\frac{1}{2i}\ln \frac{b(z)}{\tilde{b}(z)},~~|z|=1,
\\
\mathcal{P}_j=2\ln z_j,~~\mathcal{Q}_j=i\ln\gamma_j,~~1\leq j\leq N,
\\
\tilde{\mathcal{P}}_j=2\ln \tilde{z}_j,~~\tilde{\mathcal{Q}}_j=i\ln\tilde{\gamma}_j,~~1\leq j\leq \tilde{N},
\end{split}
\label{rho}
\end{eqnarray}
have the non-vanishing Poisson brackets of simple form:
\begin{eqnarray}
\begin{split}
\left\{\rho(z),\phi(\mu)\right\}=i\left(\delta(\frac{z}{\mu})+\delta(-\frac{z}{\mu})\right),
~~|z|=|\mu|=1,
\\
\left\{\mathcal{P}_j,\mathcal{Q}_k\right\}=\delta_{jk},~~1\leq j,k\leq N,
\\
\left\{\tilde{\mathcal{P}}_j,\tilde{\mathcal{Q}}_k\right\}=\delta_{jk},~~1\leq j,k\leq \tilde{N}.
\end{split}
\label{rhopb}
\end{eqnarray}

\section{Integrability of the AL system with shifted nonlocal reductions}

We are now ready to investigate the complete integrability of the AL system with local and, particularly, shifted nonlocal reductions.
We will see that the scattering data do provide the variables of action-angle type for the space shifted nonlocal DNLS equation (\ref{ssdnls}) and for the space-time shifted nonlocal DNLS equation (\ref{stsdnls}), but it is not the case for the time shifted nonlocal DNLS equation (\ref{tsdnls}) when the scattering data include  the discrete spectrum.

\subsection{Integrability of the DNLS equation}
It can be checked directly that if $\Phi(n,t,z)$ solves (\ref{LPS}) with $p(n,t)=\nu q^\ast(n,t)$, so does $\hat{\Phi}(n,t)=\sigma_{\nu}\Phi^\ast(n,t,\frac{1}{z^\ast})$, where $\sigma_\nu=\left( \begin{array}{cc} 0 & 1 \\
 \nu &  0 \\ \end{array} \right)$.
Applying this observation to the Jost solutions $T_{\pm}(n,z)$ and comparing the boundary conditions (see (\ref{JSA})), we obtain
\begin{eqnarray}
T_{\pm}^{(1)}(n,t,z)=\sigma_{\nu}\left(T_{\pm}^{(2)}(n,t,\frac{1}{z^\ast})\right)^*.
\label{Tr1}
\end{eqnarray}
Using (\ref{abexp}) and (\ref{Tr1}), we obtain the following symmetry relations for the transition coefficients:
\begin{eqnarray}
\tilde{a}(z)=a^{\ast}(\frac{1}{z^\ast}), ~~|z|\leq 1;
~~\tilde{b}(z)=\nu b^{\ast}(z),~~|z|=1.
\label{ar1}
\end{eqnarray}
It follows from (\ref{Tr1}) and (\ref{ar1}) that the eigenvalues $z_j$ and $\tilde{z}_j$ (after
rearranging) subject to
\begin{eqnarray}
\tilde{z}_j=\frac{1}{z_j^{\ast}},~~1\leq j\leq \tilde{N}=N,
\label{zsym1}
\end{eqnarray}
and the transition coefficients $\gamma_j$ and $\tilde{\gamma}_j$ subject to
\begin{eqnarray}
\tilde{\gamma}_j=\nu \gamma_j^{\ast}, ~~1\leq j\leq \tilde{N}=N.
\label{gsym1}
\end{eqnarray}

With (\ref{zsym1}) and (\ref{gsym1}) in mind, we conclude from (\ref{rho}) and (\ref{rhopb}) that the quantities
\begin{eqnarray}
\begin{split}
\rho\left(z\right)=\frac{1}{\pi}\ln\left(1+\nu |b(z)|^2\right),
~~\phi\left(z\right)=\arg b(z),~~|z|=1,
\\
p_j=-2\arg z_j,~~q_j=2\ln|\gamma_j|,~~1\leq j\leq N,
\\
\varrho_j=-2\ln |z_j|,~~\varphi_j=2\arg\gamma_j,~~1\leq j\leq N.
\end{split}
\label{aav1}
\end{eqnarray}
have the following non-vanishing Poisson brackets
\begin{eqnarray}
\begin{split}
\left\{\rho(z),\phi(\mu)\right\}=i\left(\delta(\frac{z}{\mu})+\delta(-\frac{z}{\mu})\right),
\\
\left\{p_j,q_k\right\}=\delta_{jk},
~~
\left\{\varrho_j,\varphi_k\right\}=\delta_{jk},
~~1\leq j,k\leq N.
\end{split}
\label{aap1}
\end{eqnarray}

We may represent the Hamiltonian of the DNLS equation (\ref{dnls}) as a functional of only half of the new variables (\ref{aav1}). Indeed, it follows from (\ref{Hs}), (\ref{ar1}), (\ref{zsym1}) and (\ref{aav1}) that the Hamiltonian of the DNLS equation (\ref{dnls}) can be expressed in terms of the variables  $\rho\left(\lambda\right)$, $p_j$ and $\varrho_j$ as
\begin{eqnarray}
\begin{split}
H=4\sum_{j=1}^N\varrho_j-\sum_{j=1}^N\left(e^{ip_j}+e^{-ip_j}\right)\left(e^{\varrho_j}-e^{-\varrho_j}\right)
+\frac{i}{2}\oint_{|\xi|=1}\left(\xi^{-3}+\xi-2\xi^{-1}\right)\rho(\xi)d\xi.
\end{split}
\label{H1}
\end{eqnarray}

The new variables (\ref{aav1}) completely trivialize the dynamics of the DNLS equation (\ref{dnls}). Indeed, from (\ref{aap1}) and (\ref{H1}), we have
\begin{eqnarray}
\begin{split}
\frac{d\rho\left(z\right)}{dt}=\left\{H,\rho\right\}=0,
~~\frac{d\phi\left(z\right)}{dt}=\left\{H,\phi\right\}=-\left(z-z^{-1}\right)^2,
\\
\frac{d p_j}{dt}=\left\{H,p_j\right\}=0,
~~
\frac{dq_j}{dt}=\left\{H,q_j\right\}=-i\left(e^{ip_j}-e^{-ip_j}\right)\left(e^{\varrho_j}-e^{-\varrho_j}\right), ~~j=1,\cdots, N,
\\
\frac{d \varrho_j}{dt}=\left\{H,\varrho_j\right\}=0,
~~\frac{d \varphi_j}{dt}=\left\{H,\varphi_j\right\}=4-\left(e^{ip_j}+e^{-ip_j}\right)\left(e^{\varrho_j}+e^{-\varrho_j}\right),
~~j=1,\cdots, N.
\end{split}
\label{HE1}
\end{eqnarray}

The above argument concludes the complete integrability of the DNLS equation (\ref{dnls}). The quantities (\ref{aav1}) serve as the variables of action-angle type for the DNLS equation (\ref{dnls}).

We note that the complete integrability of the DNLS equation (\ref{dnls}) has been investigated in \cite{GIK1984} by using a method, different from the $r$-matrix, which is based on expansions over the squared solutions.

\subsection{Integrability of the space shifted nonlocal DNLS equation}
Under the space shifted nonlocal reduction (\ref{nred1}), we can check directly that if $\Phi(n,t,z)$ solves (\ref{LPS}), so does $\hat{\Phi}(n,t)=\tilde{\sigma}_{\nu}\Phi^\ast(n_0-n+1,t,z^\ast)$, where $\tilde{\sigma}_\nu=\left( \begin{array}{cc} 0 & 1 \\
 -\nu &  0 \\ \end{array} \right)$.
Applying this observation to the Jost solutions $T_{\pm}(n,z)$ and comparing the boundary conditions (see (\ref{JSA})), we obtain
\begin{eqnarray}
T_{\pm}^{(2)}(n,t,z)=-\nu z^{-n_0-1}\tilde{\sigma}_{\nu}\left(T_{\mp}^{(1)}(n_0-n+1,t,z^\ast)\right)^*.
\label{Tr2}
\end{eqnarray}
These equations imply the following symmetry relations for the transition coefficients:
\begin{eqnarray}
a(z)=a^\ast(z^\ast), ~~|z|\geq 1; ~~\tilde{a}(z)=\tilde{a}^{\ast}(z^\ast), ~~|z|\leq 1;
~~\tilde{b}(z)=\nu z^{-2(n_0+1)}b^{\ast}(z^\ast),~~|z|=1.
\label{ar2}
\end{eqnarray}

The symmetry relations (\ref{ar2}) imply that the zeros of $a(z)$ and the zeros of $\tilde{a}(z)$ always appear in quartets; we denote the sets of the zeros of $a(z)$ and $\tilde{a}(z)$ by $\mathcal{Z}\equiv\left\{z_j,-z_j,z^\ast_j,-z^\ast_j\right\}_{j=1}^{N}$ and $\tilde{\mathcal{Z}}\equiv\left\{\tilde{z}_l,-\tilde{z}_l,\tilde{z}^\ast_l,-\tilde{z}^\ast_l\right\}_{l=1}^{\tilde{N}}$, respectively. The transition coefficients corresponding to $\mathcal{Z}$ becomes $\left\{\gamma_j,-\gamma_j,-\nu (z^\ast_j)^{2(n_0+1)} \frac{1}{\gamma^\ast_j},\nu (z^\ast_j)^{2(n_0+1)} \frac{1}{\gamma^\ast_j}\right\}_{j=1}^{N}$,
and the ones corresponding to $\tilde{\mathcal{Z}}$ becomes $\left\{\tilde{\gamma}_l,-\tilde{\gamma}_l,-\nu (\tilde{z}^\ast_l)^{2(n_0+1)} \frac{1}{\tilde{\gamma}^\ast_l},\nu (\tilde{z}^\ast_l)^{2(n_0+1)} \frac{1}{\tilde{\gamma}^\ast_l}\right\}_{l=1}^{\tilde{N}}$.
In terms of these scattering data, the Hamiltonian of the space shifted nonlocal DNLS equation (\ref{ssdnls})  becomes
\begin{eqnarray}
\begin{split}
H=&8\left(\sum_{j=1}^{\tilde{N}}\ln|\tilde{z}_j|-\sum_{j=1}^{N}\ln|z_j|+(N-\tilde{N})\pi i\right)
\\
&+\sum_{j=1}^N\left(z^{2}_j-z^{-2}_j+(z^\ast_j)^{2}-(z^\ast_j)^{-2}\right)
-\sum_{j=1}^{\tilde{N}}\left(\tilde{z}^{2}_j-\tilde{z}^{-2}_j+(\tilde{z}_j^\ast)^{2}-(\tilde{z}_j^\ast)^{-2}\right)
\\
&+\frac{1}{2\pi i}\oint_{|\xi|=1}\left(2\xi^{-1}-\xi^{-3}-\xi\right)
\ln\left(1+\nu \xi^{-2(n_0+1)}b(\xi)b^{\ast}(\xi^\ast)\right)d\xi.
\end{split}
\label{H2s}
\end{eqnarray}

In this situation, we introduce the quantities
\begin{eqnarray}
\begin{split}
\rho\left(z\right)=\frac{1}{\pi}\ln\left(1+\nu z^{-2(n_0+1)}b(z)b^{\ast}(z^\ast)\right),
~~\phi\left(z\right)=\frac{1}{2i}\ln \left(\nu z^{2(n_0+1)}\frac{b(z)}{b^{\ast}(z^\ast)}\right),
\\
p_j=-2\arg z_j,~~q_j=2\ln|\gamma_j|,~~\varrho_j=-2\ln |z_j|,~~\varphi_j=2\arg\gamma_j,~~1\leq j\leq N,
\\
\tilde{p}_l=-2\arg \tilde{z}_l,~~\tilde{q}_l=2\ln|\tilde{\gamma}_l|,
~~\tilde{\varrho}_l=-2\ln |\tilde{z}_l|,~~\tilde{\varphi}_l=2\arg\tilde{\gamma}_l,~~1\leq l\leq \tilde{N}.
\end{split}
\label{aav2}
\end{eqnarray}
Using (\ref{rho}) and (\ref{rhopb}), we may conclude that the above quantities have the non-vanishing Poisson brackets
\begin{eqnarray}
\begin{split}
\left\{\rho(z),\phi(\mu)\right\}=i\left(\delta(\frac{z}{\mu})+\delta(-\frac{z}{\mu})\right),
\\
\left\{p_j,q_k\right\}=\delta_{jk},
~~
\left\{\varrho_j,\varphi_k\right\}=\delta_{jk},
~~
1\leq j,k\leq N,
\\
\left\{\tilde{p}_l,\tilde{q}_m\right\}=\delta_{lm},
~~
\left\{\tilde{\varrho}_l,\tilde{\varphi}_m\right\}=\delta_{lm}
~~1\leq l,m\leq \tilde{N}.
\end{split}
\label{aap2}
\end{eqnarray}
The Hamiltonian (\ref{H2s}) can be expressed in terms of new variables (\ref{aav2}) as
\begin{eqnarray}
\begin{split}
H=&
4\left(\sum_{j=1}^N\varrho_j-\sum_{j=1}^{\tilde{N}}\tilde{\varrho}_j+2(N-\tilde{N})\pi i\right)
+\sum_{j=1}^N\left(e^{ip_j}+e^{-ip_j}\right)\left(e^{-\varrho_j}-e^{\varrho_j}\right)
\\
&-\sum_{j=1}^{\tilde{N}}\left(e^{i\tilde{p}_j}+e^{-i\tilde{p}_j}\right)\left(e^{-\tilde{\varrho}_j}-e^{\tilde{\varrho}_j}\right)
+\frac{i}{2}\oint_{|\xi|=1}\left(\xi^{-3}+\xi-2\xi^{-1}\right)\rho(\xi)d\xi.
\end{split}
\label{H2}
\end{eqnarray}
As in the standard local reduction case, only half of the variables (\ref{aav2}), namely $\rho\left(\lambda\right)$, $p_j$, $\varrho_j$, $\tilde{p}_j$ and $\tilde{\varrho}_j$, enter into the Hamiltonian (\ref{H2}).

In the new variables (\ref{aav2}), the dynamics of the space shifted nonlocal DNLS equation (\ref{ssdnls}) is completely trivialized. Indeed, from (\ref{aap2}) and (\ref{H2}), we have
\begin{eqnarray}
\begin{split}
\frac{d\rho\left(z\right)}{dt}=\left\{H,\rho\right\}=0,
~~\frac{d\phi\left(z\right)}{dt}=\left\{H,\phi\right\}=-\left(z-z^{-1}\right)^2,
\\
\frac{d p_j}{dt}=\left\{H,p_j\right\}=0,
~~
\frac{dq_j}{dt}=\left\{H,q_j\right\}=-i\left(e^{ip_j}-e^{-ip_j}\right)\left(e^{\varrho_j}-e^{-\varrho_j}\right), ~~j=1,\cdots, N,
\\
\frac{d \varrho_j}{dt}=\left\{H,\varrho_j\right\}=0,
~~\frac{d \varphi_j}{dt}=\left\{H,\varphi_j\right\}=4-\left(e^{ip_j}+e^{-ip_j}\right)\left(e^{\varrho_j}+e^{-\varrho_j}\right),
~~j=1,\cdots, N,
\\
\frac{d \tilde{p}_l}{dt}=\left\{H,\tilde{p}_l\right\}=0,
~~
\frac{d\tilde{q}_l}{dt}=\left\{H,\tilde{q}_l\right\}
=i\left(e^{i\tilde{p}_l}-e^{-i\tilde{p}_l}\right)\left(e^{\tilde{\varrho}_l}-e^{-\tilde{\varrho}_l}\right), ~~l=1,\cdots, \tilde{N},
\\
\frac{d \tilde{\varrho}_l}{dt}=\left\{H,\tilde{\varrho}_l\right\}=0,
~~\frac{d \tilde{\varphi}_l}{dt}=\left\{H,\tilde{\varphi}_l\right\}=\left(e^{i\tilde{p}_l}+e^{-i\tilde{p}_l}\right)
\left(e^{\tilde{\varrho}_l}+e^{-\tilde{\varrho}_l}\right)-4,
~~l=1,\cdots, \tilde{N}.
\end{split}
\label{HE2}
\end{eqnarray}

The above results show that (\ref{aav2}) can be interpreted as the variables of action-angle type for the space shifted nonlocal DNLS equation (\ref{ssdnls}). Thus, we obtain the complete integrability of the space shifted nonlocal DNLS model (\ref{ssdnls}) under the rapidly decreasing boundary conditions.

\subsection{Uncompatibility of the time shifted nonlocal reduction}
Under the time shifted nonlocal reduction (\ref{nred2}), we can check directly that if $\Phi(n,t,z)$ solves (\ref{LPS}), so does $\hat{\Phi}(n,t)=\sigma_{\nu}\Phi(n,t_0-t,\frac{1}{z})$, where $\sigma_\nu=\left( \begin{array}{cc} 0 & 1 \\
 \nu &  0 \\ \end{array} \right)$.
Applying this observation to the Jost solutions $T_{\pm}(n,z)$ and comparing the boundary conditions (see (\ref{JSA})), we obtain
\begin{eqnarray}
T_{\pm}^{(2)}(n,t,z)=\nu \sigma_{\nu}T_{\pm}^{(1)}(n,t_0-t,\frac{1}{z}).
\label{Tr3}
\end{eqnarray}
These equations imply the following symmetries for the transition coefficients:
\begin{eqnarray}
\tilde{a}(z)=a(\frac{1}{z}), ~~|z|\leq 1;
~~\tilde{b}(z,t)=\nu b(\frac{1}{z},t_0-t),~~|z|=1.
\label{ar3}
\end{eqnarray}
The symmetries (\ref{Tr3}) and (\ref{ar3}) imply that
\begin{eqnarray}
\tilde{z_j}=\frac{1}{z_j},~~\tilde{\gamma_j}(t)=\nu \gamma_j(t_0-t), ~~1\leq j\leq N.
\label{zred3}
\end{eqnarray}
We emphasize that this reduction, unlike the two reductions discussed above, is not compatible with the Poisson bracket relations of the scattering data in the presence of the discrete spectrum. Indeed, from the first of (\ref{gamz}) and (\ref{zred3}), we obtain $\left\{\frac{1}{z_j},\gamma_k\right\}=0$ which is in contradiction with the second of (\ref{zgamtilde}).

\subsection{Integrability of the space-time shifted nonlocal DNLS equation}
Under the space-time shifted nonlocal reduction (\ref{nred3}), we can check directly that if $\Phi(n,t,z)$ solves (\ref{LPS}), so does $\hat{\Phi}(n,t)=\tilde{\sigma}_{\nu}\Phi(n_0-n+1,t_0-t,z)$, where $\tilde{\sigma}_\nu=\left( \begin{array}{cc} 0 & 1 \\
 -\nu &  0 \\ \end{array} \right)$.
Applying this observation to the Jost solutions $T_{\pm}(n,z)$ and comparing the boundary conditions (see (\ref{JSA})), we obtain
\begin{eqnarray}
T_{\pm}^{(2)}(n,t,z)=-\nu z^{-n_0-1}\tilde{\sigma}_{\nu}T_{\mp}^{(1)}(n_0-n+1,t_0-t,z).
\label{Tr4}
\end{eqnarray}
These equations imply the following symmetry for the transition coefficients $b(z,t)$ and $\tilde{b}(z,t)$:
\begin{eqnarray}
\tilde{b}(z,t)=\nu z^{-2(n_0+1)}b(z,t_0-t),~~|z|=1.
\label{ar4}
\end{eqnarray}
Note that the space-time shifted nonlocal reduction $p(n,t)=\nu q(n_0-n,t_0-t)$, $n_0\in \mathbb{Z}[i]$, $t_0\in \mathbb{C}$, does not impose additional symmetry relations for $a(z)$ and $\tilde{a}(z)$.

Expressions (\ref{rho}), (\ref{rhopb}) and (\ref{ar4}) suggest that the variables
\begin{eqnarray}
\begin{split}
\rho\left(z\right)=\frac{1}{\pi}\ln\left(1+\nu z^{-2(n_0+1)}b(z,t)b(z,t_0-t)\right),
\\
\phi\left(z\right)=\frac{1}{2i}\ln \left(\nu z^{2(n_0+1)}\frac{b(z,t)}{b(z,t_0-t)}\right),
\\
p_j=2\ln z_j,~~q_j=i\ln\gamma_j,~~1\leq j\leq N,
\\
\tilde{p}_l=2\ln \tilde{z}_l,~~\tilde{q}_l=i\ln\tilde{\gamma}_l,~~1\leq l\leq \tilde{N},
\end{split}
\label{cvariables4}
\end{eqnarray}
have the non-vanishing Poisson brackets
\begin{eqnarray}
\begin{split}
\left\{\rho(z),\phi(\mu)\right\}=i\left(\delta(\frac{z}{\mu})+\delta(-\frac{z}{\mu})\right),
\\
\left\{p_j,q_k\right\}=\delta_{jk},~~1\leq j,k\leq N,
\\
\left\{\tilde{p}_l,\tilde{q}_m\right\}=\delta_{lm},~~1\leq l,m\leq \tilde{N}.
\end{split}
\label{cPoisson4}
\end{eqnarray}
Using (\ref{Hs}), (\ref{ar4}) and (\ref{cvariables4}), we can express the Hamiltonian of the space-time shifted nonlocal DNLS equation (\ref{stsdnls}) in terms of the new variables $\rho\left(\lambda\right)$, $p_j$ and $\tilde{p}_j$ as
\begin{eqnarray}
\begin{split}
H=&2\sum_{j=1}^{\tilde{N}}\tilde{p}_j-2\sum_{j=1}^N p_j+4(N-\tilde{N})\pi i
+\sum_{j=1}^N\left(e^{p_j}-e^{-p_j}\right)-\sum_{j=1}^{\tilde{N}}\left(e^{\tilde{p}_j}-e^{-\tilde{p}_j}\right)
\\
&+\frac{i}{2}\oint_{|\xi|=1}\left(\xi^{-3}+\xi-2\xi^{-1}\right)\rho(\xi)d\xi.
\end{split}
\label{H3}
\end{eqnarray}

From (\ref{cPoisson4}) and (\ref{H3}), we know that the new variables (\ref{cvariables4}) completely trivialize the dynamics of the space-time shifted nonlocal DNLS equation (\ref{stsdnls}):
\begin{eqnarray}
\begin{split}
\frac{d\rho\left(z\right)}{dt}=\left\{H,\rho\right\}=0,
~~\frac{d\phi\left(z\right)}{dt}=\left\{H,\phi\right\}=-\left(z-z^{-1}\right)^2,
\\
\frac{d p_j}{dt}=\left\{H,p_j\right\}=0,
~~
\frac{dq_j}{dt}=\left\{H,q_j\right\}=e^{p_j}+e^{-p_j}-2, ~~j=1,\cdots, N,
\\
\frac{d \tilde{p}_l}{dt}=\left\{H,\tilde{p}_l\right\}=0,
~~
\frac{d\tilde{q}_l}{dt}=\left\{H,\tilde{q}_l\right\}=2-e^{\tilde{p}_l}-e^{-\tilde{p}_l}, ~~l=1,\cdots, \tilde{N}.
\end{split}
\label{HE3}
\end{eqnarray}
This means that the quantities (\ref{cvariables4}) serve as the variables of action-angle type for the space-time shifted nonlocal DNLS equation (\ref{stsdnls}). Thus, the complete integrability for this equation is obtained.

\section{Integrability of the AKNS system with shifted nonlocal reductions}

In this section, we will consider another example, the well-known AKNS system
\cite{AKNS1974}
\begin{eqnarray}
\begin{array}{l}
iq_t(x,t)=q_{xx}(x,t)-2q^2(x,t)p(x,t),
\\
ip_t(x,t)=-p_{xx}(x,t)+2p^2(x,t)q(x,t).
\end{array}
 \label{akns}
\end{eqnarray}
With the Poisson brackets \cite{Faddeev2007}
\begin{eqnarray}
\left\{q(x,t),q(y,t)\right\}=\left\{p(x,t),p(y,t)\right\}=0,
~~
\left\{q(x,t),p(y,t)\right\}=i\delta(x-y),
\label{cPBakns}
\end{eqnarray}
this system can be expressed in the Hamiltonian form
\begin{eqnarray}
q_t=\left\{H,q\right\},~~p_t=\left\{H,p\right\},
\label{HEakns}
\end{eqnarray}
where
\begin{eqnarray}
\begin{split}
H=-\int_{-\infty}^{\infty}\left(q_xp_x+q^2p^2\right)dx.
\end{split}
\label{Hakns}
\end{eqnarray}
Recently, it was found in \cite{AMN2023,AM2021} that the AKNS system
admits the following new space-time shifted nonlocal reductions
\begin{eqnarray}
p(x,t)=\nu q^\ast(x_0-x,t), ~~x_0\in \mathbb{R},
\label{aknsr1}
\\
p(n,t)=\nu q(x,t_0-t), ~~t_0\in \mathbb{C},
\label{aknsr2}
\\
p(n,t)=\nu q(x_0-x,t_0-t), ~~x_0,~~ t_0\in \mathbb{C},
\label{aknsr3}
\end{eqnarray}
where $\nu=\pm 1$ and $x_0$, $t_0$ are arbitrary parameters.

\subsection{Poisson brackets of the scattering data of AKNS system}

We first recall the essential ingredients of the scattering data and the associated Poisson bracket relations for the AKNS system. It is well-known the AKNS system admits a Lax pair \cite{AKNS1974}
\begin{subequations}
\begin{eqnarray}
\phi_x(x,t,\lambda)=U(x,t,\lambda)\phi(x,t,\lambda),
~~
U(x,t,\lambda)=\left( \begin{array}{cc} -i\lambda & q \\
 r &  i\lambda \\ \end{array} \right),
 \label{lpx}
\\
\phi_t(x,t,\lambda)=V(x,t,\lambda)\phi(x,t,\lambda),
~~ V(x,t,\lambda)=\left( \begin{array}{cc}  2i\lambda^2+iqr & 2\lambda q-iq_x \\
 2\lambda r+ir_x & -2i\lambda^2-iqr  \\ \end{array} \right),
 \label{lpt}
\end{eqnarray}
\label{lpxt}
\end{subequations}
where $\lambda$ is a spectral parameter. The transition matrix $T(x,y,\lambda)$ is defined as the fundamental solution of the space-part of the Lax equations
\begin{eqnarray}
\frac{\partial T(x,y,\lambda)}{\partial x}=U(x,\lambda)T(x,y,\lambda),
\label{Tx}
\end{eqnarray}
with the initial condition $\left.T(x,y,\lambda)\right|_{x=y}=\mathbb{I}$.
Here and in the rest of this section, we have suppressed the $t$ dependence for conciseness unless there is ambiguity.
The Jost solutions $T_{\pm}(x,\lambda)$ for real $\lambda$ are defined by
\begin{eqnarray}
T_{\pm}(x,\lambda)=\lim_{y\rightarrow \pm\infty}T(x,y,\lambda)E(y,\lambda),
\label{JSakns}
\end{eqnarray}
where $E(x,\lambda)=\exp\left\{-i\lambda x \sigma_3\right\}$.
The reduced monodromy matrix $T(\lambda)$ is defined as
\begin{eqnarray}
T(\lambda)=T^{-1}_{+}(x,\lambda)T_{-}(x,\lambda).
\label{Tlam}
\end{eqnarray}
We introduce the following notation
\begin{eqnarray}
T(\lambda)=\left( \begin{array}{cc} a(\lambda) & \tilde{b}(\lambda) \\
 b(\lambda) &  \tilde{a}(\lambda)\\ \end{array} \right).
\end{eqnarray}
The analytic properties of $T_{\pm}(x,\lambda)$ imply that $a(\lambda)$ and $\tilde{a}(\lambda)$ have analytic continuation into $\im \lambda\geq 0$ and $\im \lambda\leq 0$ respectively, whereas $b(\lambda)$ and $\tilde{b}(\lambda)$ in general can only be well-defined on the real line $\im \lambda=0$. In addition, the transition coefficients satisfy
\begin{eqnarray}
a(\lambda)\tilde{a}(\lambda)-b(\lambda)\tilde{b}(\lambda)=1.
\label{abuakns}
\end{eqnarray}

We assume that $a(\lambda)$ and $\tilde{a}(\lambda)$ have only finite number of simple zeros in the upper and lower half planes, respectively. Let $\left\{\lambda_j: \im \lambda_j>0\right\}_{j=1}^{N}$ be the complete list of zeros of $a(\lambda)$, and $\left\{\tilde{\lambda}_l: \im \tilde{\lambda}_l<0\right\}_{l=1}^{\tilde{N}}$ be zeros of $\tilde{a}(\lambda)$, respectively. For $\lambda=\lambda_j$, the column $T^{(1)}_{-}(x,\lambda)$ is proportional to the column $T^{(2)}_{+}(x,\lambda)$, and for $\lambda=\tilde{\lambda}_l$, the column $T^{(2)}_{-}(x,\lambda)$ is proportional to the column $T^{(1)}_{+}(x,\lambda)$.
Let $\gamma_j$ and $\tilde{\gamma}_l$ be the proportionality coefficients,
\begin{eqnarray}
\begin{split}
T^{(1)}_{-}(x,\lambda_j)=\gamma_jT^{(2)}_{+}(x,\lambda_j),~~j=1,\cdots,N,
\\
T^{(2)}_{-}(x,\tilde{\lambda}_l)=\tilde{\gamma}_lT^{(1)}_{+}(x,\tilde{\lambda}_l),~~l=1,\cdots,\tilde{N}.
\end{split}
\label{gam}
\end{eqnarray}
It can be shown that the Hamiltonian (\ref{Hakns}) can be expressed in terms of the scattering data as
\begin{eqnarray}
\begin{split}
H=-\frac{1}{3}\left(\sum_{j=1}^{\tilde{N}}\left(2i\tilde{\lambda}_j\right)^3-\sum_{j=1}^{N}\left(2i\lambda_j\right)^3\right)
+\frac{1}{\pi}\int_{-\infty}^{\infty}\left(2i\xi\right)^2\ln\left(1+b(\xi)\tilde{b}(\xi)\right)d\xi.
\end{split}
\label{HSakns}
\end{eqnarray}

Following \cite{Faddeev2007}, we have the following Poisson bracket relations of the scattering data
\begin{subequations}
\begin{eqnarray}
\left\{\lambda_j,\lambda_k\right\}=\left\{\lambda_j,\tilde{\lambda}_l\right\}
=\left\{\tilde{\lambda}_l,\tilde{\lambda}_m\right\}
=\left\{\gamma_j,\gamma_k\right\}=\left\{\gamma_j,\tilde{\gamma}_l\right\}=\left\{\tilde{\gamma}_l,\tilde{\gamma}_m\right\}
=0,
\label{aknsp1}
\\
\left\{b(\lambda),\lambda_j\right\}=\left\{\tilde{b}(\lambda),\lambda_j\right\}=\left\{b(\lambda),\tilde{\lambda}_l\right\}
=\left\{\tilde{b}(\lambda),\tilde{\lambda}_l\right\}=0,
\label{aknsp2}
\\
\left\{b(\lambda),\gamma_j\right\}=\left\{\tilde{b}(\lambda),\gamma_j\right\}=\left\{b(\lambda),\tilde{\gamma}_l\right\}
=\left\{\tilde{b}(\lambda),\tilde{\gamma}_l\right\}=0,
\label{aknsp3}
\\
\left\{\lambda_j,\tilde{\gamma}_l\right\}=0,
~~\left\{\tilde{\lambda}_l,\gamma_j\right\}=0,
~~\left\{\lambda_j,\gamma_k\right\}=\frac{1}{2}\delta_{jk}\gamma_k,
~~\left\{\tilde{\lambda}_l,\tilde{\gamma}_m\right\}=\frac{1}{2}\delta_{lm}\tilde{\gamma}_m,
\label{aknsp4}
\\
\{b(\lambda),b(\mu)\}=\{\tilde{b}(\lambda),\tilde{b}(\mu)\}=0,
~\{b(\lambda),\tilde{b}(\mu)\}=-\pi i\delta(\lambda-\mu)\left(1+b(\lambda)\tilde{b}(\lambda)\right),
\label{aknsp5}
\end{eqnarray}
\label{abzpb}
\end{subequations}
where $1\leq j,k\leq N$, $1\leq l,m\leq \tilde{N}$.

\subsection{Integrability of the space shifted nonlocal NLS equation}
The space shifted nonlocal reduction (\ref{aknsr1}) yields the following symmetry relations for the transition coefficients \cite{AMN2023}:
\begin{eqnarray}
\begin{split}
a(\lambda)=a^\ast(-\lambda^\ast), ~~\im \lambda\geq 0; ~~\tilde{a}(\lambda)=\tilde{a}^{\ast}(-\lambda^\ast), ~~\im \lambda\leq 0;
\\
\tilde{b}(t,\lambda)=\nu e^{2i\lambda x_0}b^{\ast}(t,-\lambda),~~\lambda\in \mathbb{R}.
\end{split}
\label{aknsar1}
\end{eqnarray}
The above symmetry relations imply that the zeros of $a(\lambda)$ become $\Lambda\equiv\left\{\lambda_j,-\lambda^\ast_j: \im \lambda_j>0\right\}_{j=1}^{N}$ and the zeros of $\tilde{a}(\lambda)$ become $\tilde{\Lambda}\equiv\left\{\tilde{\lambda}_l, -\tilde{\lambda}^\ast_l: \im \tilde{\lambda}_l<0\right\}_{l=1}^{\tilde{N}}$.
The transition coefficients corresponding to the discrete spectrum $\Lambda$ becomes
$\left\{\gamma_j,-\nu e^{2i\lambda^\ast_jx_0} \frac{1}{\gamma^\ast_j}\right\}_{j=1}^{N}$,
and the ones corresponding to $\tilde{\Lambda}$ becomes
$\left\{\tilde{\gamma}_l,-\nu e^{2i\tilde{\lambda}^\ast_jx_0} \frac{1}{\tilde{\gamma}^\ast_l}\right\}_{l=1}^{\tilde{N}}$.
With these scattering data, we introduce the quantities
\begin{eqnarray}
\begin{split}
\rho\left(\lambda\right)=\frac{1}{2\pi}\ln\left(1+\nu e^{2i\lambda x_0}b(\lambda)b^{\ast}(-\lambda)\right),
~~\phi\left(\lambda\right)=i\ln \left(\nu e^{2i\lambda x_0}\frac{b^{\ast}(-\lambda)}{b(\lambda)}\right),
\\
p_j=2\re \lambda_j,~~q_j=2\ln|\gamma_j|,~~\varrho_j=-2\im \lambda_j,~~\varphi_j=2\arg\gamma_j,~~1\leq j\leq N,
\\
\tilde{p}_l=2\re \tilde{\lambda}_l,~~\tilde{q}_l=2\ln|\tilde{\gamma}_l|,
~~\tilde{\varrho}_l=-2\im \tilde{\lambda}_l,~~\tilde{\varphi}_l=2\arg\tilde{\gamma}_l,~~1\leq l\leq \tilde{N},
\end{split}
\label{aakns1}
\end{eqnarray}
which have the non-vanishing Poisson brackets 
\begin{eqnarray}
\begin{split}
\left\{\rho(\lambda),\phi(\mu)\right\}=\delta(\lambda-\mu),
\\
\left\{p_j,q_k\right\}=\delta_{jk},
~~
\left\{\varrho_j,\varphi_k\right\}=\delta_{jk},
~~
1\leq j,k\leq N,
\\
\left\{\tilde{p}_l,\tilde{q}_m\right\}=\delta_{lm},
~~
\left\{\tilde{\varrho}_l,\tilde{\varphi}_m\right\}=\delta_{lm}
~~1\leq l,m\leq \tilde{N}.
\end{split}
\label{aapakns1}
\end{eqnarray}

Proceeding as in section 4.2, we find the Hamiltonian of the space shifted nonlocal NLS equation can be expressed in terms of new variables (\ref{aakns1}) as
\begin{eqnarray}
\begin{split}
H=\frac{2}{3}\left(\sum_{j=1}^N\left(\varrho^3_j-3p^2_j\varrho_j\right)
-\sum_{j=1}^{\tilde{N}}\left(\tilde{\varrho}^3_j-3\tilde{p}^2_j\tilde{\varrho}_j\right)\right)
-8\int_{-\infty}^{\infty}\xi^{2}\rho(\xi)d\xi.
\end{split}
\label{Hakns1}
\end{eqnarray}
Under the new variables (\ref{aakns1}), the dynamics of the space shifted nonlocal NLS equation become:
\begin{eqnarray}
\begin{split}
\frac{d\rho\left(\lambda\right)}{dt}=\left\{H,\rho\left(\lambda\right)\right\}=0,
~~\frac{d\phi\left(\lambda\right)}{dt}=\left\{H,\phi\left(\lambda\right)\right\}=-8\lambda^2,
\\
\frac{d p_j}{dt}=\left\{H,p_j\right\}=0,
~~
\frac{dq_j}{dt}=\left\{H,q_j\right\}=-4 p_j\varrho_j, ~~j=1,\cdots, N,
\\
\frac{d \varrho_j}{dt}=\left\{H,\varrho_j\right\}=0,
~~\frac{d \varphi_j}{dt}=\left\{H,\varphi_j\right\}=2\left(\varrho_j^2-p^2_j\right),
~~j=1,\cdots, N,
\\
\frac{d \tilde{p}_l}{dt}=\left\{H,\tilde{p}_l\right\}=0,
~~
\frac{d\tilde{q}_l}{dt}=\left\{H,\tilde{q}_l\right\}
=4 \tilde{p}_l\tilde{\varrho}_l, ~~l=1,\cdots, \tilde{N},
\\
\frac{d \tilde{\varrho}_l}{dt}=\left\{H,\tilde{\varrho}_l\right\}=0,
~~\frac{d \tilde{\varphi}_l}{dt}=\left\{H,\tilde{\varphi}_l\right\}=2\left(\tilde{p}^2_l-\tilde{\varrho}_l^2\right),
~~l=1,\cdots, \tilde{N}.
\end{split}
\label{HEakns1}
\end{eqnarray}

The above results mean that the new variables (\ref{aakns1}) serve as the canonical variables of action-angle type for the space shifted nonlocal NLS equation, and thus the complete integrability of the space shifted nonlocal NLS equation is obtained.

\subsection{Uncompatibility of the time shifted nonlocal reduction}
Under the time shifted nonlocal reduction (\ref{aknsr2}), the transition coefficients $a(k)$ and $\tilde{a}(k)$ subject to the symmetry \cite{AMN2023}
\begin{eqnarray}
\tilde{a}(\lambda)=a(-\lambda),~~\tilde{b}(t,\lambda)=\nu b(t_0-t,-\lambda).
\label{aknsar2}
\end{eqnarray}
This implies that the discrete spectrum and the corresponding transition coefficients (after rearranging) subject to
\begin{eqnarray}
\tilde{\lambda}_j=-\lambda_j,~~\tilde{\gamma_j}(t)=\nu \gamma_j(t_0-t), ~~1\leq j\leq N.
\label{zred3}
\end{eqnarray}
Such a restriction is not compatible with the Poisson bracket relations (\ref{aknsp4}). This means that, as in the AL case, the time shifted nonlocal reduction is not compatible with the
Poisson bracket relations of the scattering data in the presence of the discrete spectrum.

\subsection{Integrability of the space-time shifted nonlocal DNLS equation}
The space-time shifted nonlocal reduction (\ref{aknsr3}) implies the following symmetry for the transition coefficients $b(\lambda,t)$ and $\tilde{b}(\lambda,t)$:
\begin{eqnarray}
\tilde{b}(\lambda,t)=\nu e^{2i\lambda x_0}b(\lambda,t_0-t),~~\lambda\in \mathbb{R}.
\label{aknsar3}
\end{eqnarray}
Note that, as in the AL case, the space-time shifted nonlocal reduction (\ref{aknsr2}) does not impose additional symmetry relations for $a(\lambda)$ and $\tilde{a}(\lambda)$.

In this case, we introduce
\begin{eqnarray}
\begin{split}
\rho\left(\lambda\right)=\frac{1}{2\pi}\ln\left(1+\nu e^{2i\lambda x_0}b(\lambda,t)b(\lambda,t_0-t)\right),
~~\phi\left(\lambda\right)=i\ln \left(\nu e^{2i\lambda x_0}\frac{b(\lambda,t_0-t)}{b(\lambda,t)}\right),
\\
p_j=2\lambda_j,~~q_j=\ln\gamma_j,
~~
\tilde{p}_l=2 \tilde{\lambda}_l,
~~\tilde{q}_l=\ln\tilde{\gamma}_l,
~~1\leq j\leq N,~~1\leq l\leq \tilde{N}.
\end{split}
\label{aakns2}
\end{eqnarray}
These variables have the non-vanishing Poisson brackets
\begin{eqnarray}
\begin{split}
\left\{\rho(\lambda),\phi(\mu)\right\}=\delta\left(\lambda-\mu\right),
\\
\left\{p_j,q_k\right\}=\delta_{jk},~~1\leq j,k\leq N,
\\
\left\{\tilde{p}_l,\tilde{q}_m\right\}=\delta_{lm},~~1\leq l,m\leq \tilde{N}.
\end{split}
\label{aapakns2}
\end{eqnarray}

The Hamiltonian in this case can be expressed in terms of new variables (\ref{aakns2}) as
\begin{eqnarray}
\begin{split}
H=\frac{i}{3}\sum_{j=1}^{\tilde{N}}\tilde{p}^3_j-\frac{i}{3}\sum_{j=1}^Np^3_j
-8\int_{-\infty}^{\infty}\xi^{2}\rho(\xi)d\xi.
\end{split}
\label{Hakns2}
\end{eqnarray}
As in the previous cases, the new variables (\ref{aakns2}) play the role of canonical variables of action-angle type for the space-time shifted nonlocal NLS equation; they completely trivialise the dynamics of the space-time shifted nonlocal NLS equation:
\begin{eqnarray}
\begin{split}
\frac{d\rho\left(\lambda\right)}{dt}=\left\{H,\rho\left(\lambda\right)\right\}=0,
~~\frac{d\phi\left(\lambda\right)}{dt}=\left\{H,\phi\left(\lambda\right)\right\}=-8\lambda^2,
\\
\frac{d p_j}{dt}=\left\{H,p_j\right\}=0,
~~
\frac{dq_j}{dt}=\left\{H,q_j\right\}=-i p^2_j, ~~j=1,\cdots, N,
\\
\frac{d \tilde{p}_l}{dt}=\left\{H,\tilde{p}_l\right\}=0,
~~
\frac{d\tilde{q}_l}{dt}=\left\{H,\tilde{q}_l\right\}
=i \tilde{p}^2_l, ~~l=1,\cdots, \tilde{N}.
\end{split}
\label{HEakns2}
\end{eqnarray}

\section{Concluding remarks}

This paper aims to study the complete integrability of soliton equations with space or/and time shifted nonlocal reductions in the sense of Liouville (in the infinite-dimensional Hamiltonian setting). We investigated the construction of the variables of action-angle type and the Hamiltonian descriptions in terms of these variables for the space or/and time shifted nonlocal systems. We illustrated our results in detail on two examples: one is the typical semi-discrete model, the AL system, and the other is the well-known continuous model, the AKNS system. We believe that a similar argument will work for other space-time shifted nonlocal soliton models, such as the space-time shifted nonlocal mKdV equation and the multi-components NLS equations with space-time shifted nonlocal reductions.

\section*{ACKNOWLEDGMENTS}

This work was supported by the National Natural Science Foundation of China (Grant No. 12271221).

\vspace{1cm}
\small{

}

\end{CJK}
\end{document}